\documentclass[12pt]{article}
\usepackage{epsf}
\makeatletter

\def\ltap{\raisebox{-.4ex}{\rlap{$\sim$}} \raisebox{.4ex}{$<$}}
\def\gtap{\raisebox{-.4ex}{\rlap{$\sim$}} \raisebox{.4ex}{$>$}}

\begin{document}
\begin{titlepage}
\today          \hfill
\begin{center}
\hfill    LBNL-41338 \\
\hfill    UCB-PTH-97/65 \\
\vskip .25in
{\large \bf Supersymmetry Breaking 
and the Supersymmetric Flavour Problem: An Analysis 
of Decoupling the First Two Generation Scalars 
}
\footnote{This work was supported in part
by the Director, Office of Energy
Research, Office of High Energy and Nuclear Physics,
Division of High
Energy Physics of the U.S. Department of Energy
under Contract
DE-AC03-76SF00098. 
KA was also supported
by the Berkeley Graduate Fellowship and
MG by NSERC.}

\vskip .25in

\vskip .25in
K. Agashe\footnote{email: ksagashe@lbl.gov.} and M. Graesser
\footnote{email: mlgraesser@lbl.gov}

{\em Theoretical Physics Group\\
    Lawrence Berkeley National Laboratory\\
    University of California, Berkeley, California 94720
and \\
    Department of Physics \\
      University of California, Berkeley, California 94720} 
\end{center}

\vskip .25in
\begin{abstract}
The supersymmetric contributions to
the Flavor Changing
Neutral Current processes may be
suppressed by decoupling the
scalars of the first and second generations.
It is known, however, that the heavy scalars drive the stop
mass squareds negative through the two-loop
Renormalization Group 
evolution.
This tension is studied in detail.
Two new items are included in this analysis:
the effect of
the top quark Yukawa coupling and the QCD
corrections to the
supersymmetric contributions to $\Delta m_K$.
Even with Cabibbo-like 
degeneracy between the
squarks of the first two generations, these squarks
must be heavier than $\sim 40$ TeV
to suppress $\Delta m_K$. This implies, in the 
case of a high scale of supersymmetry breaking, that 
the boundary value of the stop mass has to be
greater than                               
$\sim 7$ TeV to keep the stop mass squared 
positive at the weak scale. Low-energy supersymmetry breaking
at a scale that is of the same order as the
mass of the heavy scalars is also considered. 
In this case the finite
parts of
the two-loop diagrams are computed to estimate the
contribution of the heavy scalar masses to the
stop mass squared. It is found that for 
Cabibbo-like mixing between the squarks, 
the stop mass
at the boundary needs to be larger than $\sim 2$ TeV. 
Thus, for both cases, the large boundary value 
of the stop masses leads to  
an unnatural amount of fine tuning to obtain the correct
$Z$ mass.

\end{abstract}
\end{titlepage}
\renewcommand{\thepage}{\roman{page}}
\setcounter{page}{2}
\mbox{ }

\vskip 1in

\begin{center}
{\bf Disclaimer}
\end{center}

\vskip .2in

\begin{scriptsize}
\begin{quotation}
This document was prepared as an account of work sponsored by the
United
States Government. While this document is believed to contain
correct
 information, neither the United States Government nor any agency
thereof, nor The Regents of the University of California, nor any
of their
employees, makes any warranty, express or implied, or assumes any legal
liability or responsibility for the accuracy, completeness,
or usefulness
of any information, apparatus, product, or process disclosed, or
represents that its use would not infringe privately owned rights.
Reference herein
to any specific commercial products process, or service by
its trade name,
trademark, manufacturer, or otherwise, does not necessarily
constitute or
imply its endorsement, recommendation, or favoring by the
United States
Government or any agency thereof, or The Regents of the
University of
California.  The views and opinions of authors expressed herein
do not
necessarily state or reflect those of the United States
Government or any
agency thereof, or The Regents of the University of California.
\end{quotation}
\end{scriptsize}

\vskip 2in

\begin{center}
\begin{small}
{\it Lawrence Berkeley National Laboratory is an equal opportunity employer.}
\end{small}
\end{center}

\newpage
\renewcommand{\thepage}{\arabic{page}}
\setcounter{page}{1}

\section{Introduction}

The origin of electroweak symmetry breaking (EWSB) and the subsequent 
gauge hierarchy problem \cite{susskind} are 
two large mysteries of 
the Standard Model (SM).  
Supersymmetry (SUSY)\cite{fayet} provides a promising solution to these 
problems, by both stabilising the weak scale against 
radiative corrections\cite{nonrenorm}, and by naturally breaking the 
electroweak symmetry through the quantum corrections of the superpartner 
of the top 
quark to the Higgs boson 
mass \cite{ross}.  
 It is known, however, that generic weak scale values for the 
masses of the first two generation 
scalars give rates for many flavour violating processes
that are in disagreement with the experimental observation. The measured 
value of $\Delta m_K$ and detection limits for
$\mu\rightarrow e \gamma$, and $\mu\rightarrow3 e$, 
 for example, require that the first two generation 
scalars be degenerate to within a few tenths of a percent if their
masses are at the weak scale \cite{susyflpblm,fcnc}. Constraints from
$CP$ violation are generally even more severe. 
Understanding the origin of this degeneracy is the supersymmetric flavour 
problem. Attempts to resolve this puzzle  
without introducing any fine tuning 
include: using approximate
non-abelian or abelian symmetries\cite{hall}; 
communicating supersymmetry breaking to the visible sector 
by gauge interactions that 
do not distinguish between flavours \cite{gm}; squark-quark 
mass matrix alignment \cite{seiberg}; and  
raising the soft 
masses of the first two generation scalars 
to the tens of TeV range \cite{dine,pomoral,dvali,nelson2,nelson3,nelson,mohapatra,dimopoulos}.
 
The phenomenological viability and naturalness of this 
last scenario is the subject of this
paper. To suppress flavour changing processes, the heavy scalars must 
have masses between a  
few TeV and a hundred TeV. 
The actual value depends on the degree of degeneracy and mixing 
between the masses of the 
first two generation scalars. 
As discussed in Reference \cite{nima}, the masses of 
the heavy scalars cannot be made arbitrarily large without 
breaking colour and charge. 
This is because 
the heavy scalar masses contribute to the two-loop 
Renormalisation 
Group Equation (RGE) for the soft masses of the light scalars, such
that 
the stop soft mass squared become 
more negative in RG scaling 
to smaller energy scales. This negative contribution is large if 
the scale at which supersymmetry breaking is communicated
to the visible sector is close to the Grand Unification scale\cite{nima}.
With the first two generation soft scalar masses 
$\approx$ 10 TeV, the initial value of the 
soft masses for the light stops must be $\approx 
(\hbox{few} $ TeV$)^2$ 
to cancel this negative contribution \cite{nima} to  
obtain the correct vaccum. This requires, however, an unnatural amount
of fine tuning to correctly break the electroweak 
symmetry\cite{finetune,anderson}.

In this paper we analyse these issues and include two new items not previously
discussed within this context:
the effect of the large top quark Yukawa coupling, $\lambda_t$,
in the RG evolution, that drives the stop soft mass squared more negative;
and QCD
radiative corrections in the $\Delta m_K$ 
constraint \cite{bagger}. 
This modifies the bound on the heavy scalar masses which is 
consistent with the measured value of $\Delta m_K$. This, in 
turn, affects the minimum value of the initial scalar masses that is 
required to keep the scalar masses positive at the weak scale. 

We note that the severe constraint obtained for the initial 
stop masses
assumes that supersymmetry breaking occurs at a high scale.
This 
leaves open the possibility that requiring positivity 
of the scalar masses is not a strong constraint if the 
scale of supersymmetry breaking is not much larger than the 
mass scale of the heavy scalars. In this paper 
we investigate this possibility 
by computing the finite parts of the same two-loop diagrams
responsible for the negative contribution to the light scalar 
RG equation, and use these results as an {\em estimate} 
of the two-loop contribution in an actual model of low-energy 
supersymmetry breaking.
We find that in certain 
classes of models, requiring positivity of the soft masses 
may place strong necessary conditions that such models must 
satisfy in order to be phenomenologically viable.

Our paper is organized as follows. In Section 2 an overview of the 
ingredients of our analysis is presented. Some philosophy and notation is
discussed. Section 2.1 discusses the constraints on the masses and 
mixings of the first two generation scalars obtained from $\Delta m_K$
after including QCD corrections. It is found, in particular, 
 that Cabibbo-like mixing 
among both the first two generation left-handed squarks 
and right-handed squarks 
requires them to be heavier than 40 TeV. Section 2.2 discusses the 
logic of our RG analysis, and some formulas are presented. 
This analysis is independent of the $\Delta m_K$ analysis. 
Sections 3 and 
4 apply this machinery to the cases of low-energy and high-energy supersymmetry
breaking, respectively.  
Section 3 deals with the case in which the scale at 
which SUSY
breaking is communicated to the SM sparticles 
is close to
the mass of the heavy scalars. We use the finite parts 
of the two-loop
diagrams to estimate the negative contribution of the 
heavy scalars.
We find that Cabibbo-like mixing among the left-handed and right-handed
first two generation squarks 
implies that the boundary value of the stop masses 
has to greater
than $\sim 2$ TeV to keep
the stop mass squareds positive at the
weak scale. 
This results in a fine tuning of
naively $1\%$ in electroweak symmetry breaking \cite{finetune}. 
We also discuss the cases where 
there is $O(1)$ mixing among only the right or left
squarks of the first two generations, 
and find that requiring 
positivity of the
slepton mass squareds implies a constraint 
on the stop masses of $\sim 1$ TeV 
if gauge-mediated boundary conditions 
are used to relate the two masses. This is comparable to the 
direct constraint on the initial stop masses. 
In Section 4, we consider the case where the SUSY 
breaking masses
for the SM sparticles are generated at a high scale 
($\sim 10^{16}$ GeV).
In this case, the negative contribution of the 
heavy scalars
is enhanced by a large logarithm. 
We consider various boundary conditions for the
stop and Higgs masses and find that with $O(0.22)$ 
degeneracy
between the first two generation squarks, the 
boundary value
of the stop mass needs to be larger than $\sim 7$ TeV.
This gives a fine tuning of naively 
$0.02 \%$\cite{finetune}. 
For $O(1)$ mixing
between the left (right) squarks only, the minimum 
initial value
of the stop is $\sim 4 (2)$ TeV. In Section 5 the scale of 
supersymmetry breaking is varied between 50 TeV and 
$2 \times 10^{16}$ GeV. Uppers bounds on the amount of 
degneracy required between the first two generation scalars, that is 
consistent with positivity of the light scalar masses, naturalness 
in electroweak symmetry breaking, and the measured value of 
$\Delta m_K$, are obtained. 
These results are summarized in Figures \ref{m2ftd1} and \ref{m2ftd2}.
We conclude in Section 6. In the Appendix, we discuss 
the computation
of the two-loop diagrams which give the negative 
contribution of 
the heavy scalars to the light scalar mass squareds.

\section{\bf Overview.}
\label{setup}
The chiral 
particle content of the Minimal Supersymmetric Standard
Model (MSSM) contains 3 generations of ${\bf \bar{5}}$+${\bf 10}$
representations of $SU(5)$. The supersymmetry must be softly
broken to not be excluded by experiment. 
Thus the theory must also be supplemented by some `bare' soft
supersymmetry breaking parameters, as well as a physical 
cutoff, $M_{SUSY}$. 
The
`bare' soft supersymmetry breaking parameters are then the
coefficients appearing in the Lagrangian, defined with a
cutoff $M_{SUSY}$. It will be assumed for simplicity 
that the bare soft masses,
$\tilde{m}^2_{i,0}$, the bare gaugino masses $M_{A,0}$, and a bare
trilinear term for the stops, $\lambda_t A_{t,0}$, are all 
generated close to this scale. 
The MSSM is then a good effective theory at energies 
below the scale $M_{SUSY}$,
but above the mass 
of the heaviest superpartner.

The physical observables at low-energies will depend on these
parameters.
If an unnatural degree of cancellation is required between the bare 
parameters of the theory to produce a measured observable, the theory
may be considered 
to be fine tuned. 
Of course, it is possible that a more fundamental theory
may resolve in a natural manner 
the apparent 
fine tuning.  
The gauge-hierarchy problem is a well-known example of this. 
The Higgs boson mass
of the SM
is fine tuned if the SM is valid at energies above a few TeV.
This fine tuning is removed if at energies close to the weak scale 
the SM is replaced by a more fundamental 
theory that is supersymmetric\cite{nonrenorm}. 
 
One quantification of the fine tuning of an observable ${\cal O}$
with respect to a bare parameter $\lambda_0$  
is given by Barbieri-Giudice \cite{finetune} to be 
\begin{equation}
\Delta({\cal O};\lambda_0)
=(\delta {\cal O}/ {\cal O})/(\delta \lambda_0 / \lambda_0)
=\frac{\lambda_0}{{\cal O}}\frac{\partial}{\partial \lambda_0} {\cal O}.
\label{ft}
\end{equation}
It is argued that this only measures the sensitivity of ${\cal O}$ to 
$\lambda_0$, and care should be taken when interpreting whether 
a large value of $\Delta$ necessarily implies that ${\cal O}$ is fine tuned
\cite{anderson}. It is not the intent of this paper to quantify fine tuning; 
rather, an estimate of the fine tuning is sufficient and Equation 
\ref{ft} will be used. In this paper the value of 
${\cal O}$ is considered extremely unnatural if $\Delta({\cal O}; 
\lambda_0) >100$.

The theoretical prediction for $\Delta m_K$ (within the MSSM) 
and its measured value 
are an example of such a fine tuning: 
Why should the masses of the first two generation scalars be degenerate 
to within 1 GeV, when their masses are $O(\hbox{500 GeV})$? Phrased 
differently, the first two generation scalars must be extremely 
degenerate for the MSSM to not be excluded by the measured value 
of $\Delta m_K$. 
An important direction in supersymmetry model building is aimed 
at attempting to explain the origin of this degeneracy.

One proposed solution to avoid this fine tuning 
is to decouple the first two generation scalars 
since their masses are the most stringently constrained by the 
flavour violating processes \cite{dine,pomoral,
dvali,nelson2,nelson,mohapatra,dimopoulos}. In 
this scenario, some of the first two generation 
scalars have masses $M_S \gg m_Z$. 
To introduce some notation,
$n_5$ $(n_{10})$ will denote the number of ${\bf \bar{5}}$ $({\bf 10})$
scalars of the MSSM particle content
that are very heavy \footnote{It is assumed that the heavy scalars 
form complete $SU(5)$ multiplets to avoid a large Fayet-Illiopoulus $D-$
term at the one-loop level\cite{dimopoulos,nelson2}.}.
Thus at energy scales $E \ll M_S$ the particle
content is that of the MSSM, minus the $n_5$ ${\bf \bar{5}}$ and $n_{10}$ {\bf 10}
scalars.
In the literature this is often
referred to as `The More Minimal Supersymmetric
Standard Model'\cite{nelson2}.  

There are, however, other possible and {\it equally valid} 
sources of fine tunings. The measured 
value of the $Z$ mass is such an example \cite{finetune}. 
The minimum of the renormalized
Higgs potential determines the value of the $Z$ mass which is already known 
from experiment. The vacuum expectation value (vev) of 
the Higgs field is, in turn, a function of 
the bare parameters of the theory. The relation used here, valid at the 
tree-level, is 
\begin{equation}
\frac{1}{2}m^2_Z = - \mu^2
+ \frac{m^2_{H_d}(\mu_G)-m^2_{H_u}(\mu_G) \tan^2 \beta}
{\tan^2 \beta-1}.
\label{muEW}
\end{equation} 
It is clear from this Equation that requiring 
correct electroweak symmetry breaking 
relates the value of the soft Higgs masses at the weak scale, 
$m^2_{H_d}(\mu_G)$ and $m^2_{H_u}(\mu_G)$,  
to the supersymmetric Higgs mass $\mu$.
A numerical computation determines the dependence of
$m^2_{H_u}(\mu_G)$ and $m^2_{H_d}(\mu_G)$ 
on the bare parameters 
$M_{A,0}$, $m^2_{\tilde{t}_i,0}$ and $M_S$. 
In the MSSM, the cancellation required 
between
the bare parameters of the theory for it not to be 
excluded by the $Z$ mass 
increases as the scale of supersymmetry 
breaking is increased. Typically, the bare mass of the gluino, stops, 
and the first two generation
squarks must  
be less than a few TeV and ten TeV, respectively, 
for successful electroweak symmetry breaking not to be 
fine tuned at more than the one per cent level \cite{finetune,anderson,
dimopoulos}.

These two potential 
fine tuning problems- the supersymmetric flavour problem and that 
of electroweak symmetry breaking- are not completely independent, for they 
both relate to the size of supersymmetry breaking \cite{dimopoulos,nima}. 
Thus the consistency of  
any theoretical framework that attempts to resolve one fine tuning 
issue can be tested by requiring that it not
reintroduce any comparable 
fine tunings in other sectors of the theory. 
This is the situation for the case under consideration 
here. Raising the masses of the first two generation scalars 
can resolve the supersymmetric flavour problem. 
As discussed in \cite{dimopoulos}, this results in
a fine tuning of $m^2_Z$ through the two-loop dependence of 
$m^2_{H_u}(\mu_G)$ on $M_S$.
There is, however, another source of fine tuning of
$m_Z$ due to the heavy scalars:
these large masses require that the bare masses 
of the stops, in particular, be typically larger than 
a few TeV to keep the soft 
masses squared positive at the weak scale \cite{nima}. This large value for 
the bare stop mass 
prefers a large value for the vev of the Higgs field, thus 
introducing a fine tuning 
in the electroweak sector. Further, this fine tuning is typically 
not less than the 
original fine tuning in the flavour sector.
This is the central issue of this paper.

\subsection{$\Delta m_K$ Constraints}
\label{mk}
At the one-loop level the 
exchange of gluinos and squarks generates a 
$\Delta S = 2$ operator. In the limit
$M_3 << M_S$
that we are interested in, the
$\Delta S = 2$ effective Lagrangian at the scale
 $M_S$ obtained by 
integrating out
the squarks
is
\begin{equation}
{\cal L}_{eff} = \frac{ \alpha _S ^2 (M_S) }{216 M_S^2} 
\left(C_1 {\cal O} _1 +
\tilde{C}_1 \tilde{ {\cal O} }_1 +
C_4 {\cal O} _4 + C_5 {\cal O} _5+\hbox{h.c.}\right).
\end{equation}
Terms that are $O(M^2_3 /M^2_S)$ are subdominant and 
neglected. We  
expand the exact result in powers of $\delta_{LL,RR}=s_{L,R} c_{L,R} \eta_{L,R} (\tilde{m}^2_1-
\tilde{m}^2_2)_{LL,RR}/
\tilde{m}^2_{AV,L,R}$, where $\tilde{m}^2_{AV}$ is the average mass 
of the scalars, and where $\eta_{L,R}$ is the phase 
and $s_{L,R}$ is the $1-2$ 
element of the $W_{L,R}$ matrix that appears at the 
gluino-squark-quark vertex\footnote{In this paper 
only 1-2 generation mixing is considered. Direct $L-R$ mass mixing 
is also neglected.}. This 
approximation underestimates the magnitude of the 
exact result, so our analysis is conservative\cite{nima}. 
The coefficients $C_i$ to 
leading order in $\delta_{LL}$, $\delta_{RR}$, are
\begin{eqnarray}
C_1 & = & -22 \delta^d _{LL} \nonumber \\
C_4 & = & 24 \delta^d _{LL} \delta^d _{RR} 
\nonumber \\
C_5 & = & -40 \delta^d _{LL} \delta^d _{RR}.
\end{eqnarray}
The coefficient $\tilde{C}_1$ is obtained from $C_1$ with 
the replacement $\delta^d_{LL} \rightarrow \delta^d_{RR}$.
The operators ${\cal O}_i$ are
\begin{eqnarray}
{\cal O}_1 & = & \bar{d} ^a _L \gamma _{\mu}s_{L,a} 
\bar{d} ^b _L \gamma ^{\mu}
s_{L,b}
\nonumber \\
{\cal O}_4 & = & \bar{d} ^a _R s_{L,a} \bar{d} ^b _L s_{R,b} 
\nonumber \\
{\cal O}_5 & = & \bar{d} ^a _R s_{L,b} \bar{d} ^b _L s_{R,a}
\end{eqnarray}
and $\tilde{ {\cal O} }_1$ is obtained from ${\cal O}_1$
with the replacement $L \rightarrow R$.
The Wilson coefficients, $C_1 - C_5$, are RG scaled 
from the scale of the
squarks, $M_S$ , to 900 MeV
using the anomalous dimensions of the operators, ${\cal O}_1 
- {\cal O}_5$.
The anomalous dimension of ${\cal O}_1$ is well known
\cite{gaillard} and is $\mu d C_1 /d\mu =\alpha_s C_1/  \pi $.
We have
computed the other anomalous dimensions
and our result agrees with that of \cite{bagger} (see this
reference for a more general analysis
of QCD corrections to the SUSY contributions to $K -\bar{K}$
mixing). 
These authors , however, choose to RG scale to $\mu_{had}$, 
defined by $\alpha_s(\mu_{had})$=1. The validity 
of the pertubation expansion  
is questionable at this scale; we choose instead 
to RG scale to 900 MeV, where
$\alpha_s(\hbox{900 MeV}) \sim .4$. 
The result is
\begin{eqnarray}
C_1 (\mu _{had}) & = & \kappa _1 C_1 (M_S) 
\nonumber \\
\tilde{C}_1 (\mu _{had}) & = & \kappa _1 \tilde{C}_1 (M_S) 
\nonumber \\
C_4 (\mu _{had}) & = & \kappa _4 C_4 (M_S) + \frac{1}{3}
(\kappa _4 - \kappa _5) C_5 (M_S) \nonumber \\
C_5 (\mu _{had}) & = & \kappa _5 C_5 (M_S)
\end{eqnarray}
where
\begin{eqnarray}
\kappa _1 & = & \left( \frac{\alpha _s (m_c)}
{\alpha _s (\hbox{900 MeV})} \right)^{6/27}
\left( \frac{\alpha _s (m_b)
}{\alpha _s (m_c)} \right)^{6/25} \left( \frac{\alpha _s (m_t)}
{\alpha _s (m_b)} \right)^{6/23}
 \left( \frac{\alpha _s
(\mu _G)}{\alpha _s (m_t)} \right)^{6/21} 
\left( \frac{\alpha _s (M_S)}
{\alpha _s
(\mu _G)} \right)^{ 6/( 9 + (n_5 + 3 n_{10})/2 ) } \nonumber \\
\kappa _4 & = & \kappa _1 ^{-4} \nonumber \\
\kappa _5 & = & \kappa _1 ^{1/2}
\end{eqnarray}
The effective Lagrangian at the hadronic 
scale is then
\begin{equation}
{\cal L} _{eff} = \frac{ \alpha _s ^2 (M_S) }{216 M_S^2}
\left( -22 (\delta^d_{LL})^2 \kappa _1 {\cal O}_1 - 22 
(\delta^d_{RR})^2 \kappa _1
\tilde{{\cal O}}_1
 + \delta^d _{LL} \delta^d _{RR} ( \frac{8}{3} 
(4 \kappa _4 + 5 \kappa _5)
 {\cal O}_4 - 40 \kappa _5 {\cal O}_5 )+\hbox{h.c.}\right). 
\end{equation}
The SUSY contribution to the $K - \bar{K}$ mass difference is
\begin{equation}
(\Delta m_K)_{SUSY} = 2 \hbox{Re} < K | {\cal L}_{eff} | \bar{K} >.
\end{equation}
The relevant matrix elements (with bag factors set to 1) are
\begin{eqnarray}
< K | {\cal O}_1 | \bar{K} > & = & \frac{1}{3} m_K f^2_K 
\nonumber \\
< K | {\cal O}_4| \bar{K} > & = & \left ( \frac{1}{24} 
+ \frac{1}{4} \left(
\frac{m_K}{m_s + m_d} \right)^2 \right ) m_K f^2_K \nonumber \\
< K | {\cal O}_5 | \bar{K} > & = & \left ( \frac{1}{8} + 
\frac{1}{12} \left(
\frac{m_K}{m_s + m_d} \right)^2 \right ) m_K f^2_K
\end{eqnarray}
in the vacuum insertion approximation. We use \cite{pdg} 
$m_K =497$ MeV, $f_K =160$ MeV, $m_s =150$ MeV
, $(\Delta m_K)_{exp}=3.5 \times 10^{-12}$ MeV, and 
$\alpha _s (M_Z) = 0.118$. This gives $\alpha_s(m_b)=.21$, 
$\alpha_s(m_c)=.29$ and $\alpha_s(\hbox{900 MeV})=.38$ using 
the one-loop RG evolution. 
Once values for $(n_5,n_{10},\delta^d_{LL},\delta^d_{RR})$ are
specified, a minimum value for $M_S$ is gotten by requiring that 
$(\Delta m_K)_{SUSY} =(\Delta m_K)_{exp}$.
In the case that both $\delta_{RR}\neq0$ and $\delta_{LL} \neq 0$, 
we assume that both the left-handed and right-handed squarks are
heavy, so that $(n_5,n_{10})=(2,2)$. 
In this case we require that only the
dominant contribution to $\Delta m_K $, which
is $\sim \delta^d_{LL} \delta^d_{RR}$,
equals the measured value of $\Delta m_K$. 
If $\delta_{RR} \neq0$ and $\delta_{LL}=0$, we assume that 
only the right-handed squarks are heavy, and thus $(n_5,n_{10})=(2,0)$.
Similarly, if $\delta_{LL} \neq0$ and $\delta_{RR}=0$ then 
$(n_5,n_{10})=(0,2)$.
Limits are given in Tables 1 and 2 for some
choices of these parameters. These results agree with Reference 
\cite{bagger} for the same choice of input parameters. 
For comparison, the limits gotten by 
neglecting the QCD corrections are also 
presented in Tables 1 and 2.
We consider $\delta^d_{LL}$ $(\delta^d_{RR})=
(i)$ $1$, $(ii)$ $.22$, $(iii)$ $.1$, and $(iv)$ $0.04$.
These correspond to :$(i)$ no mixing and no degeneracy;
$(ii)$ Cabibbo-like mixing; $(iii)$ Cabibbo-like mixing and
$\sim .5$ degeneracy; and $(iv)$ Cabibbo-like mixing and 
Cabibbo-like degeneracy.
We expect only cases $(i)$, $(ii)$ and $(iii)$ to be  
relevant if the 
supersymmetric flavour problem is resolved by 
decoupling the first two generation
scalars.  
From Table \ref{mktable} we note that for $(n_5,n_{10})=
(2,0)$, $M_S$ must 
be larger than $\sim$ 30 TeV if it is assumed there is 
no small mixing or degeneracy  
$(\delta^d_{RR}=1)$ between the first two generation 
scalars. 

\begin{table}
\begin{center}
\begin{tabular}{||l|l|l||} \hline
$\sqrt{\hbox{Re}(\delta^d_{LL}\delta^d_{RR})}$ & 
$(n_5,n_{10})=(2,2)$ & $(n_5,n_{10})=(2,2)$\\ \hline
 & QCD incl. & no QCD \\ \hline
$1$ & $\hbox{182 TeV}$ & $\hbox{66 TeV}$ \\ \hline
$.22$ & $\hbox{40 TeV}$ & $\hbox{15 TeV}$ \\ \hline
$.1$ & $\hbox{18 TeV}$ & $\hbox{7.3 TeV}$ \\ \hline
$0.04$ & $\hbox{7.3 TeV}$ & $\hbox{3.1 TeV}$  \\ \hline
\end{tabular}
\end{center}
\caption{Minimum values for heavy scalar masses 
$M_S$ obtained from
the measured value of $\Delta m_K$ assuming 
$M^2_3/M^2_S\ll 1$. The limits labeled `QCD incl.' include 
QCD corrections as discussed in the text. Those labeled as `no QCD' 
do not.}
\protect\label{mktable0}
\end{table}

\begin{table}
\begin{center}
\begin{tabular}{||l|l|l||}\hline
$\hbox{Re}(\delta^d_{RR}) \hbox{ }(\delta^d_{LL}=0)$ 
& $(n_5,n_{10})=(2,0)$ & $(n_5,n_{10})=(2,0)$\\
 \hline
& QCD incl. & no QCD \\ \hline
$1$ & $\hbox{30 TeV}$ & $\hbox{38 TeV}$ \\ \hline
$.22$ & $\hbox{7.2 TeV}$ & $\hbox{8.9 TeV}$ \\ \hline
$0.1$ & $\hbox{3.4 TeV}$ & $\hbox{4.1 TeV}$\\ \hline
$0.04$ & $\hbox{1.4 TeV}$ & $\hbox{1.7 TeV}$ \\ \hline
\end{tabular}
\end{center}
\caption{Minimum values for heavy scalar masses
$M_S$ obtained from
the measured value of $\Delta m_K$ assuming 
$M^2_3/M^2_S \ll 1$. The limits labeled as `QCD incl.' 
include QCD 
corrections as discussed in the text. Those labeled as `no QCD'
do not. 
The limits for
$(n_5,n_{10})=(0,2)$ obtained by $\delta_{LL}^d 
\leftrightarrow \delta_{RR}^d$ are similar 
and not shown.}
\label{mktable}
\end{table}

The limits gotten from the mesaured rate of $CP$ violation are now briefly
discussed.
Recall that the $CP$ violating parameter $\epsilon$ is approximately
\begin{equation}
|\epsilon| \sim \frac{|\hbox{Im}<K|{\cal{L}}_{eff}|\bar{K}>|}
{\sqrt{2} \Delta m_K},
\label{cp}
\end{equation}   
and its measured value is 
$|\epsilon| \sim |\eta_{00}|
=$2.3$\times$10$^{-3}$ \cite{pdg}. In this case, the 
small value of $\epsilon$ implies either that the 
phases appearing in the soft scalar mass matrix 
are extremely tiny, or that the masses of the heavy scalars are 
larger than the limits given in Tables \ref{mktable0} 
and \ref{mktable}. In the case where the phases are $O(1)$, 
$\hbox{Im}<K|{\cal{L}}_{eff}|\bar{K}> \sim 
\hbox{Re}<K|{\cal{L}}_{eff}|\bar{K}>
$ and thus the stronger constraint on $M_S$ is 
obtained from $\epsilon$ and not $\Delta m_K$, 
for the same choice of input parameters. 
In particular, the constraint from $CP$ violation 
increases the minimum allowed value of $M_S$ by a 
factor of $1/ \sqrt{2 \sqrt{2} \epsilon} \sim$12.5. 
This significantly increases the minimum value of the 
initial light scalar masses that is allowed by the positivity requirement. 
   
\subsection{RGE analysis}
The values of the soft masses at the weak scale are determined by
the RG evolution.
In the $\overline{DR}'$ scheme \cite{dred,epscalar,drbarp}, the RG
equations \footnote{An earlier version of the analysis presented in 
this manuscript did not 
include the one-loop hypercharge $D-$term in the RG equations. This has  
been corrected. This has not changed our conclusions though, since the 
numerical effect of this error is small. The authors 
thank B. Nelson for drawing our attention to this omission.} for the 
light scalar masses are, including the
gaugino, $A$-term and $\lambda_t$ contributions at
the one-loop level and the heavy scalar
contribution at the two-loop level \cite{twolooprge},
\begin{eqnarray}
\frac{d}{dt}m_{i}^2(t=\ln \mu)&=& -\frac{2}{\pi}\sum_{A} \alpha_A(t)
C_A^i M^2_A(t)
+\frac{4}{16 {\pi}^2} \sum_{A} C_A^i \alpha^2_A(t)
(n_5 m^2_5+3 n_{10} m^2_{10}) \nonumber \\
& &+\frac{8}{16 {\pi}^2}\frac{3}{5} Y_{i} \alpha_1(t)
\left(\frac{4}{3}
\alpha_3(t)-\frac{3}{4}\alpha_2(t)-\frac{1}{12} \alpha_1(t)\right)
(n_5 m^2_5-
n_{10} m^2_{10}) \nonumber \\
& & +\frac{\eta_i \lambda_t^2(t)}{8 {\pi} ^2}
(m^2_{H_u} (t)+m^2_{\tilde{u}^c_3} (t) +
m^2_{\tilde{Q}_3}(t)+A(t)^2) \nonumber \\
& & + \frac{6}{5} \frac{1}{4 \pi} Y_i \alpha_1(t) \hbox{Tr} Ym^2(t),
\label{rg1}
\end{eqnarray}
with $\eta=(3,2,1)$ for $\tilde{f}_i =H_u$, $\tilde{t}^c$,
$\tilde{t}$, respectively, and zero otherwise.
For simplicity it is assumed that
$M_{A,0}/\alpha_{A,0}$ are all equal at $M_{SUSY}$.
The initial value of the gluino mass, $M_{3,0}$, is then chosen to be
the independent parameter.
To avoid a large Fayet-Illiopoulus $D$-term at
the one-loop level, we assume that the heavy scalars form complete
$SU(5)$ representations\cite{dimopoulos,nelson2}.
We use $SU(5)$ normalisation for the $U(1)$ coupling constant and
$Q=T_3+Y$. Finally,
$C_A^i$ is the quadratic Caismir for the gauge group
$G_A$ that is
$4/3$ and $3/4$ for the fundamental representations
of $SU(3)$ and $SU(2)$, and $3/5Y^2_i$
for the $U(1)$ group.
The cases
$(n_5,n_{10})$= (I) $(2,2)$, (II) $(2,0)$, (III) $(0,2)$
are considered.
The results for the case $(3,0)$ is obtained, to a good approximation,
 from Case (II) by a
simple scaling, and it is not discussed any further.

Inspection of Equation $($\ref{rg1}$)$
reveals that in RG scaling from a high scale to a smaller scale
the two-loop gauge contribution to the soft masses is negative, and
that of the gauginos is positive.
The presence of the
large $\lambda_t$ Yukawa coupling in the RGE drives the value
of the stop soft mass squared even more
negative. This effect increases the
bound on the initial value for the stop
soft masses and is included in our analysis.
In our analysis the top quark mass in $\overline{MS}$ scheme 
is fixed at 167 GeV.

In the MSSM there is an extra parameter, $\tan\beta$, which is the
ratio of the vacuum expecations values of the Higgs fields that
couple to the up-type and down-type quarks respectively.
Electroweak symmetry breaking then
determines the top quark mass to be
$m_t=\lambda_t/\sqrt{2} v\sin \beta$ with $v\sim$ 247 GeV.
In our analysis we consider the regime of small to moderate
$\tan\beta$, so that all Yukawa couplings other than $\lambda_t$
are neglected in the RG evolution.
In this approximation
the numerical results for $\tilde{f}_i\neq \tilde{t}$ or
$\tilde{t}^c$ are independent of $\tan\beta$.
In the numerical analysis of Sections \ref{lowsusy} and \ref{highsusy} 
$\tan\beta$=2.2 is considered.
In Section \ref{sectionft} $\tan \beta=10$ is also considered.

In the case of low-energy supersymmetry breaking,
the scale $M_{SUSY}$
is not much larger than
the mass scale of the heavy scalars. Then the
logarithm $\sim$$\ln(M_{SUSY}/M_S)$ that
appears in the
solution to the previous RG equations is only $O(1)$.
In this case the finite parts of the
two-loop diagrams may not be negligible and should be included in
our analysis. We use these finite parts to {\em estimate} the size
of the two-loop heavy scalar contribution in an actual model.

The full-two loop expression for the soft scalar mass at a
renormalisation scale $\mu_R$ is
$m^2_{full}(\mu_R)=m^2_{\overline{DR}'}(\mu_R)+m^2_{finite}(\mu_R)$, where
$m^2_{\overline{DR}'}(\mu_R)$ is the solution to the RG equation in
$\overline{DR}'$ scheme, and $m^2_{finite}(\mu_R)$ is the finite part
of the one-loop and
two-loop diagrams, also computed in $\overline{DR}'$ scheme.
The finite
parts of the two-loop diagrams that contain internal heavy scalars 
are computed in the Appendix and the details are given therein.
The answer for these two-loop finite parts is 
(assuming all heavy scalars are degenerate with common
mass $M_S^2$)
\begin{eqnarray}
m^2_{i,finite}(\mu_R)&=&-\frac{1}{8}(\ln (4 \pi)-\gamma
+ \frac{\pi^2}{3}-2-
\ln\left(\frac{M^2_S}{\mu_R^2})\right)
\times \sum_{A}
{\left(\frac{\alpha_A(\mu_R)}{\pi}\right)}^2 (n_5+3 n_{10}) C_A^i
 M^2_S \nonumber \\
& &-\frac{3}{5}\frac{1}{16 {\pi}^2}
 \alpha_1(\mu_R)(n_5-n_{10}) Y_i
\left(6-\frac{2}{3} {\pi}^2+2 (\ln (4 \pi)-\gamma)
-4 \ln(\frac{M_S^2}{\mu_R^2}) \right)
\nonumber \\
& & \times
\left(\frac{4}{3} \alpha_3(\mu_R)-\frac{3}{4}
\alpha_2(\mu_R)-\frac{1}{12}\alpha_1(\mu_R)\right) M_S^2
\label{finite}
\end{eqnarray}
where the gaugino and fermion masses are neglected.
Since we use the $\overline{DR}^{\prime}$ scheme to compute 
the finite parts of the soft scalar masses, the limits we obtain
on the initial masses 
are only valid, strictly speaking, in this scheme. This is 
especially relevant for the case of low scale SUSY breaking. 
So 
while these
finite parts should be viewed as semi-quantitative, they should
suffice for a discussion of the fine tuning that results from the 
limit on the bare stop mass.
For the case of high scale SUSY breaking, the RG logarithm
is large and so the finite parts are not that important.

Our numerical analysis for either low-energy or
high-energy supersymmetry breaking is described as follows.

The RG equations are evolved from the scale $M_{SUSY}$
to the scale at which the heavy scalars are decoupled. 
This
scale is denoted by $\mu_S$ and should be $O(M_S)$. 
The RG scaling  
of the heavy scalars is neglected.
At this scale the finite parts of the two-loop diagrams
are added to $m^2_{\tilde{f}_i}(\mu_S)$. We note that since the two-loop
information included in our RG analysis is the leading $O(M^2_S)$ effect,
it is sufficient to only use tree-level matching at the scale
$\mu_S$. Since the heavy scalars are not
included in the effective theory
below $M_S$ and do not contribute to the gauge coupling beta
functions, the numerical results
contain an
implicit dependence
on the number of heavy scalars. This results in a smaller value for
$\alpha_3(\mu_S)$ compared to its value if instead all the scalars have a $\sim 1$TeV mass. This tends to weaken the constraint, and so it
is included in our analysis \footnote{This is the origin of a small 
numerical discrepancy of $\sim 10\%$ between our results and the 
analysis of \cite{nima} in the approximation $\lambda_t=0$.}.
The soft masses are then evolved using the one-loop
RGE to the mass scale at which the
gluinos are decoupled. This scale is fixed to be $\mu_G$=1 TeV. 

A constraint on the initial value of the soft
masses is obtained by requiring 
that at the weak scale the physical
scalar masses are positive.
The experimental limit is $\sim$ 70 GeV for charged or 
coloured scalars\cite{LEP}. The physical mass of a scalar is equal 
to the sum of  
the soft scalar mass, the electro-weak $D-$term, the supersymmetric 
contribution, and 
some finite one-loop and two-loop contributions. As mentioned in the previous paragraph, in the effective theory below
$M_S$ the finite two-loop part from the heavy scalars is included in 
value of the 
soft scalar mass of the light sparticles 
at the boundary, defined at $\mu_R =\mu_S \sim M_S$.
The finite one-loop  
contributions
are proportional to the gaugino and other light scalar masses, and are smaller
than the corresponding logarithm that is summed in $\tilde{m}^2_i(\mu_R)$. So
we neglect these finite one-loop parts. Further, the electroweak $D-$terms 
are less than 70 GeV. For the scalars other than the stops, the 
supersymmetric contribution is negligible. 
In what follows then, we will require that 
$\tilde{m}^2_i(\mu_G)>0$
for scalars other than the 
stops.
The discussion with the stops is complicated by both the large
supersymmetric contribution, $m^2_t$, to the physical mass and
by the $L-R$ mixing between the gauge eigenstates. This mixing
results
in a state with mass squared less than
$\hbox{min}(m^2_{\tilde{t}}+m^2_t,
m^2_{\tilde{t}^c}+m^2_t)$, so it is a conservative
assumption to
require that for both gauge eigenstates the value of
$m^2_{\tilde{t}_{i}}+m^2_t$ is larger than the
experimental limit. 
This implies that 
$m^2_{\tilde{t}_{i}}\gtap ($70 GeV$)^2-$$($175 GeV$)^2
=-($160 GeV$)^2$.
In what follows we require instead that
$m^2_{\tilde{t}_{i}}\geq 0$.
This results in an error that is
$ (160 $GeV$)^2/2m_{\tilde{t}_{i,0}}
\approx 26 $ GeV if
the constraint obtained by neglecting $m_t$ is $\sim$ 1 TeV.
For the parameter range of interest it will be shown that the
limit on the initial squark masses
is $\sim$ 1 TeV, so this approximation is consistent.

We then combine the above two analyses as follows.
The $\Delta m_K$ constraints of Section \ref{mk}
determine a minimum value for $M_S$
once some theoretical preference for
the $\delta$'s is given.
Either a natural
value for the $\delta$'s is predicted by some model, 
or the $\delta$'s are
arbitrary and chosen solely by naturalness considerations.
Namely, in the latter case the fine tuning to suppress 
$\Delta m_K$ is roughly $2/\delta$. 
Further, a model may also predict the ratio $M_3 /M_S$. 
Otherwise, Equations \ref{ft} and \ref{muEW} may be used as a rough
guide to determine an upper value for $M_3$, based upon naturalness
considerations of the $Z$ mass.
Without such a limitation, the
positivity requirements are completely irrelevant if the 
bare gluino mass is suffuciently large; 
but then the $Z$ mass is fine tuned.
Using these values of $M_3$ and $M_S$, 
the RGE analysis gives a minimum value for the initial
stop masses which is consistent with $\Delta m_K$ and 
positivity of the soft masses.
This translates into some fine tuning of the $Z$ mass,
which is then roughly quantified by
Equations \ref{ft} and \ref{muEW}.

Finally, we remark that our analysis may also be extended to include models 
that contain a Fayet-Illiopoulos hypercharge $D-$term, $\zeta_D$, 
at the tree-level. The effect of the $D-$term is to shift the soft scalar 
masses, $m^2_{i,0} \rightarrow \tilde{m}^2_{i,0}=m^2_{i,0} +
Y_i \zeta_D$. 
In this case, the positivity analysis applies to $\tilde{m}^2_{i,0}$, rather 
than $m^2_{i,0}$.

\section{Low Energy Supersymmetry Breaking}
\label{lowsusy}
 In this Section we investigate the 
positivity requirement within a 
framework that satisifes both of the 
following: (i)
supersymmetry breaking is communicated to the visible sector 
at low energies; and (ii) multi-TeV scale soft masses, $M_S$, 
are generated for some of the first two generation scalars.
This differs from the usual low-energy supersymmetry breaking 
scenario in that we assume $M^2_S \gg m^2_{\tilde{t}_i,0}$. 
In the absence of a specific model, however, 
it is difficult to obtain from the positivity 
criterion 
robust constraints on the 
scalar spectra for the following reasons. 
At the scale $M_{SUSY}$ it is expected that, 
in addition to the heavy scalars of the MSSM, there are 
particles
that may have SM quantum numbers and supersymmetry 
breaking mass parameters. All these extra states contribute 
to the soft scalar masses of the light particles. 
The sign of this
contribution depends on, among other things,
whether the soft mass squared for these
additional particles is positive or negative-clearly very 
model-dependent. The total two-loop 
contribution to the light scalar 
masses is thus a sum of a model-dependent
part and a model independent part.
By considering only the model-independent 
contribution we have only isolated one particular contribution
to the total value of the soft scalar masses near the supersymmetry
breaking scale. We will, however, use these results to 
{\em estimate}  
the typical size of the finite parts in an actual model. 
That is, if in an actual model 
the sign of the finite parts is negative and 
its size
is of the same magnitude as in
 Equation $($\ref{finite}$)$, the constraint in that model 
is identical
to the constraint that we obtain. The constraint for other values 
for the finite parts is then obtained from our results by a simple scaling.

Before discussing the numerical results, the 
size of the finite contributions are estimated
in order to
illustrate the problem. Substituting $M_S\sim$ 25
TeV, $\alpha_3($25 TeV$)\sim$ 0.07 
and $\alpha_1($25 TeV$)\sim$ 0.018 into Equation \ref{finite}
gives
\begin{equation}
\delta m^2_{\tilde{q}}\approx-(\hbox{410 GeV})^2(n_5+3 n_{10})
\left(\frac{M_S}{\hbox{25 TeV}}\right)^2
\end{equation}
for squarks, and
\begin{equation}
\delta m^2_{\tilde{e}^c}\approx
-\left((n_5+3 n_{10})
(\hbox{70 GeV})^2
+(n_5-n_{10})(\hbox{100 GeV})^2\right)
\left(\frac{M_S}{\hbox{25 TeV}}\right)^2
\end{equation}
for the right-handed selectron. The negative contribution is
large if $M_S\sim$ 25 TeV. For example,
if $n_5=n_{10}=2$ then $\delta m^2_{\tilde{e}^c}\approx
-($200 GeV$)^2$ and $\delta m^2_{\tilde{q}}\approx-($1.2 TeV$)^2$.
If $n_5=2$, $n_{10}=0$, then $\delta m^2_{\tilde{e}^c}\approx
-($170 GeV$)^2$ and $\delta m^2_{\tilde{q}}\approx-($580 GeV$)^2$.

In this low-energy supersymmetry breaking scenario, it is 
expected that $M_{SUSY}\sim M_S$. In our numerical 
analysis we will set $M_{SUSY}=\mu_S$ since the actual
messenger scale is not known. The scale $\mu_S$ is chosen 
to be 50 TeV. At the scale $\mu_S=$50 TeV the 
$\mu _S$-independent parts
of Equation $($\ref{finite}$)$ are added to the initial 
value of the soft scalar masses.  
The soft masses are then 
evolved using the RG equations (not including the two-loop
contribution) to the scale $\mu_G$= 1TeV.

\begin{figure}
\centerline{\epsfxsize=0.5\textwidth \epsfbox{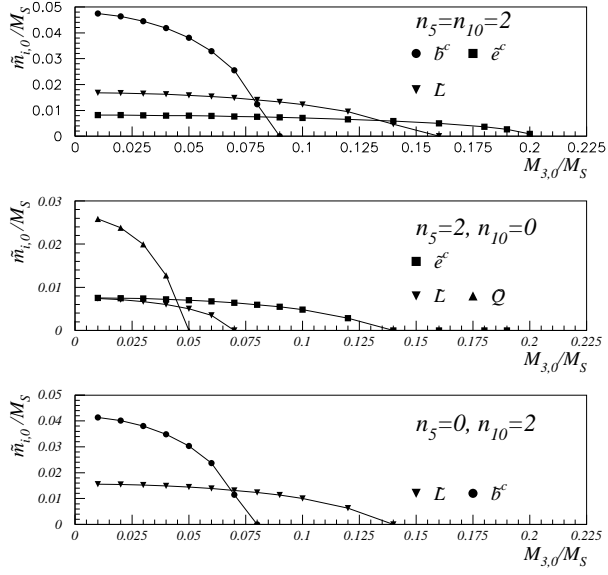}} 
\caption{Limits for $m_{\tilde{f}_i,0}/M_S$ 
from the requirement that the mass squareds are
positive at the weak scale, for low-energy 
supersymmetry breaking. The regions below the curves
are excluded. For the case (2,0),
the limits for the other squarks are very similar to
that for $\tilde{Q}$ and are therefore not shown.}
\protect\label{m2l}
\end{figure}

First we discuss the constraints the positivity requirement 
imply for $\tilde{f}_i\neq \tilde{t}_L$ or $\tilde{t}_R$.
In this case $m^2_{\tilde{f}_i}$ is
renormalised by
$M^2_{3,0}$, $M^2_S$, $m^2_{\tilde{f}_i,0}$ and the initial 
value of $\hbox{Tr} Y m^2 \equiv D_{Y,0}$. We find
\begin{eqnarray}
m^2_{\tilde{f}_i}(\mu_G) & = & m^2_{\tilde{f}_i,0}
+(0.243 C_3^i+0.0168 C_2^i+0.00156 Y^2_i) M^2_{3,0}
+c_D \times 10^{-3} Y_i D_{Y,0} \nonumber \\
& & -(0.468 C_3^i+.095 C_2^i+.0173 Y^2_i)
\frac{1}{2}(n_5+3 n_{10})\times 10^{-3} M^2_S \nonumber \\
& & -0.0174 (n_5-n_{10}) Y_i \times 10^{-3} M^2_S \nonumber \\
& & -(n_5-n_{10})\left((-0.00058+0.0016(n_5+3n_{10})) M^2_S 
-.925 M^2_{3,0}\right) Y_i  \times 10^{-3}, 
\label{m2}
\end{eqnarray}
where the strongest dependence on $(n_5,n_{10})$ has been 
isolated. The coefficient appearing in front 
of $D_{Y,0} $ is $c_D =-6.$ 
The numerical coefficients in Equation 
(\ref{m2}) also depend on $(n_5,n_{10})$ and the numbers presented
in Equation(\ref{m2}) are for $(n_5,n_{10})=($2,0$)$. This 
sensitivity is, however, only a few percent between the 
four cases under consideration here \footnote{This dependence 
is included in Figure \ref{m2l}.}. 
Requiring
positivity of the soft scalar masses directly constrains
$m^2_{\tilde{f}_i,0}/M^2_S$ and $M^2_{3,0}/M^2_S$.

The value of $D_{Y,0}$ depends on the spectrum at the supersymmetry 
breaking scale, and is therefore model-dependent. To obtain model-independent 
constraints from the positivity requirement, we therefore only constrain 
the combination $\tilde{m}^2_{\tilde{f}_i,0} \equiv 
m^2_{\tilde{f}_i,0}+c_D Y_i D_{Y,0}$.  
Only this combination appears in the weak-scale 
value for the scalar mass of $\tilde{f}_i$. 
The numerical effect is small, since 
with $D_{Y,0} \sim O(m^2_{\tilde{f}_i,0})$, the coefficient 
of $m^2_{\tilde{f}_i,0}$ 
is shifted from $1$ to $\sim 1-6. \times 10^{-3} Y_i$.

The positivity requirement $\tilde{m}^2_{\tilde{f}_i}$ for
$\tilde{f}_i\neq \tilde{t}$ or $\tilde{t}^c$ is given
in Figure \ref{m2l} for different values of $n_5$ and $n_{10}$.
That is, in Figure \ref{m2l}  
the minimum value of $\tilde{m}_{\tilde{f}_i,0}/M_S$ required to 
keep the soft masses positive at the scale $\mu_G$ is plotted 
versus $M_{3,0}/M_S$.
We conclude from these figures that the positivity criterion is
weakest for $n_5$=2 and $n_{10}$=0. This is expected since in
this case the heavy particle content is the smallest. We note 
that even
in this `most minimal' scenario the negative contribution to the masses
are rather large. In particular, we infer from Figure \ref{m2l} 
that for $(n_5=2,n_{10}=0)$ and
$M_S\sim$ 25 TeV, $\delta m^2_{\tilde{e}^c} \approx-($190 GeV$)^2$
for $M_{3,0}$ as large as 1 TeV.
In this case it is the two-loop contribution from the
hypercharge $D$-term
that is responsible for the large negative mass squared.
In the case $(n_5,n_{10})$=$(2,2)$, 
we obtain from Figure \ref{m2l} that
for $M_S\sim$ 25 TeV, 
$\delta m^2_{\tilde{e}^c} \approx-($210 GeV$)^2$
and $\delta m^2_{\tilde{b}^c} \approx-($1.1 TeV$)^2$ for 
$M_{3,0}$ as large as 1 TeV.

We now apply the positivity requirement to the stop sector. 
In this case it is not possible to directly constrain the
boundary values of the stops for the following simple reason.
There are only two positivity constraints, whereas the
values of $m^2_{\tilde{t}}(\mu_G)$
and $m^2_{\tilde{t}^c}(\mu_G)$ are functions of the three
soft scalar masses $m^2_{\tilde{t},0}$,
$m^2_{\tilde{t}^c,0}$ and $m^2_{H_u,0}$. To obtain a
limit some theoretical assumptions must be made to relate
the three initial soft scalar masses.

The numerical solutions to the RG equations 
for $\tan\beta$=2.2 and $(n_5,n_{10})=(2,0)$ are
\begin{eqnarray}
m^2_{\tilde{t}}(\mu_G) & = & -0.0303 A_t^2+ 0.00997 A_t M_{3,0}
+ 0.322 M^2_{3,0} + c_D \times \frac{1}{6} \times 10^{-3} D_{Y,0}  \nonumber \\
& & - 0.0399 (m^2_{H_u,0}+m^2_{\tilde{t}^c,0})
+ 0.960 m^2_{\tilde{t},0} - 0.000645 c_{L} M^2_S \nonumber  \\
m^2_{\tilde{t}^c}(\mu_G) & = & -0.0606 A_t^2  + 0.0199 A_t M_{3,0} 
+ 0.296 M^2_{3,0}  +c_D \times \frac{-2}{3} \times 10^{-3} D_{Y,0} \nonumber  \\
& & 0.920 m^2_{\tilde{t}^c,0}- 0.0797 (m^2_{H_u,0}+m^2_{\tilde{t},0})
- 0.000492 c_{R} M^2_S \nonumber \\
m^2_{H_u}(\mu_G) & = & -0.0909 A_t^2  + 0.0299 A_t M_{3,0} 
- 0.0289 M^2_{3,0}  +c_D \times \frac{1}{2} \times 10^{-3} D_{Y,0} \nonumber \\
& & + 0.880 m^2_{H_u,0} -0.119 (m^2_{\tilde{t},0}
+m^2_{\tilde{t}^c,0})+0.0000719 c_{H} M^2_S.
\label{stopsol}
\end{eqnarray}
The numerical coefficients other than that of $M_S$ do not 
vary more than a few percent between the 
different values for $(n_5,n_{10})$, and thus this dependence is not shown. 
For $M_S$, we find that  $(c_L,c_R,c_H)$ is 
$(1,1,1)$, $(3.62,3.84,4.59)$, 
$(2.78,3.04,3.92)$, for $(n_5,n_{10})=($2$,$0$)$, $(2,2)$
and $(0,2)$, respectively. Also, $c_D=-6.$ 
We find from Equations \ref{ft} and \ref{muEW}  
that to keep $m^2_Z$ fine tuned
at less than $1\%$ ($\Delta \leq 100$) in each of the bare parameters,
we must have:
$\mu \ltap$ 460 GeV; $M_{3,0} \ltap 2.3 
$ TeV; $m_{\tilde{t},0} \ltap 1.7$ TeV; $m_5 \ltap 
80$ TeV and $m_{10} \ltap 50$ TeV for $(n_5,n_{10})=(2,2)$. 
Finally, for other values of these parameters
the fine tuning increases as
$\Delta= 100 \times \tilde{m}^2/ \tilde{m}^2_0$,
where $\tilde{m}_0$ is the value
of $\tilde{m}$ that gives $\Delta=100$.

It is possible to show, using the fact that 
$Y_{H_u}+Y_{Q}+Y_{u^c}=0$, that the solutions in Equations 
\ref{stopsol} are unchanged if we replace $m^2_{i,0}$ with 
$\tilde{m}^2_{i,0}=m^2_{i,0}+ c_D \times 10^{-3} Y_i D_{Y,0}$ {\it and} 
set $D_{Y,0}=0$. 
In what follows then, we will use the posivitity analysis to constrain 
$\tilde{m}^2_{i,0}$ for the stops. We note though, that the difference between 
$\tilde{m}^2_{i,0}$ and 
$m^2_{i,0}$ is small, 
owing to the small coefficient appearing in front of  
$D_{Y,0}$. In the remainder of this Section the tilde on $\tilde{m}^2_{i,0}$ 
will be removed to simplify the notation. 
 
\begin{figure}
\centerline{\epsfxsize=0.5\textwidth \epsfbox{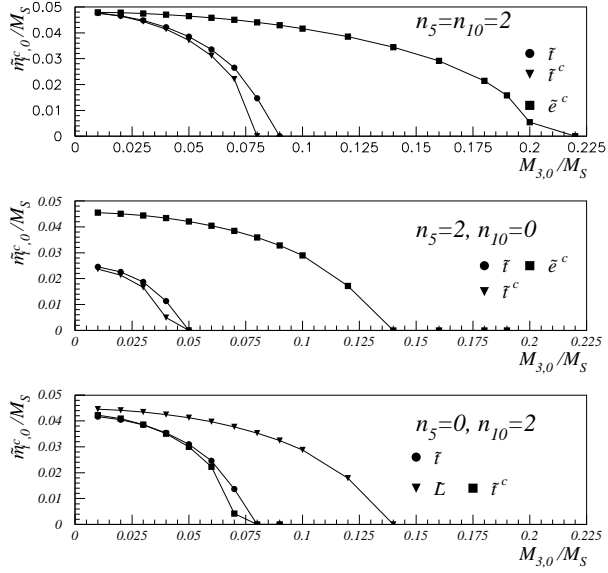}}
\caption{Limits for $m_{\tilde{t}^c,0}/M_S$
from the requirement that the stop and slepton mass squared are
positive at the weak scale.
The regions below the curves
are excluded.
Low-energy 
gauge-mediated
supersymmetry breaking mass relations between the light sparticles 
and $\tan \beta=$2.2 
are assumed.}
\protect\label{m2lgm}
\end{figure}

To constrain the initial values of the stop masses
 we will only consider gauge-mediated supersymmetry breaking 
mass relations. From Equation \ref{stopsol} we see that to naturally 
break electroweak symmetry a small hierarchy 
$m^2_{\tilde{t}_i,0} > m^2_{H_u,0}$ is required. 
This is naturally 
provided by gauge-mediated boundary conditions \footnote{In fact, low-energy
gauge-mediated supersymmetry breaking provides ``too
much'' electroweak symmetry breaking \cite{gmfinetune}.}.  
The relations between the soft scalar masses when
supersymmetry breaking is communicated to the visible 
sector by gauge messengers
are \cite{gm} 
\begin{equation}
m^2_{i,0}
=\frac{3}{4}\sum_{A}C_A^i
\frac{\alpha^2_A(M_{SUSY})}
{\alpha^2_3(M_{SUSY})+\alpha^2_1(M_{SUSY})/5}
 m^2_{\tilde{t}^c,0}.
\label{GMBC}
\end{equation}
Substituting these relations into Equations (\ref{stopsol})
and assuming $A_{t,0}=$0 
determines $m^2_{\tilde{t}}(\mu_G)$ and 
$m^2_{\tilde{t}^c}(\mu_G)$ as a function of $M_{3,0}$, $M^2_S$ 
and $m^2_{\tilde{t}^c,0}$. In Figure \ref{m2lgm} we 
have plotted the minimum value of $m_{\tilde{t}^c,0}/M_{3,0}$ 
required to maintain both $m^2_{\tilde{t}}(\mu_G)\geq0$ and
$m^2_{\tilde{t}^c}(\mu_G)\geq0$. 

Another interesting constraint on these class
of models is found if it is assumed that the 
initial masses of all the 
light fields are related at the supersymmetry breaking scale 
by some gauge-mediated supersymmetry breaking (GMSB) mass relations, as
in Equation (\ref{GMBC}). This ensures
the degeneracy, as required by the flavour changing constraints, of any 
light scalars of the first two generations. This is required if, 
for example, one of 
$n_5$ or $n_{10}$ are zero.  
Then in our previous limits of 
$m_{\tilde{f}_i,0}$ for $\tilde{f}_i\neq\tilde{t}$ or 
$\tilde{t}^c$, constraints on the initial 
value of $m_{\tilde{t}^c}$ are obtained 
by relating $m_{\tilde{f}_i,0}$ 
to  $m_{\tilde{t}^c,0}$ using Equation (\ref{GMBC}). 
In this case the slepton masses provide the 
strongest constraint and they are also shown in Figure 
\ref{m2lgm}.
This result may be understood from the 
following considerations.
The two-loop hypercharge $D$-term 
contribution to the soft mass 
is  $\sim Y_i (n_5-n_{10}) \alpha_1 \alpha_3 M^2_S$
and this has two interesting consequences. The first is that
for $n_5 \neq n_{10}$, 
the resulting $\delta \tilde{m}^2$ is 
always negative for one of $\tilde{e}^c$ or $\tilde{L}$. 
Thus in this case there 
is always a constraint on $m^2_{\tilde{t}^c}$ 
once gauge-mediated boundary conditions are assumed.
That this negative contribution is large is 
seen as follows. The combined tree-level mass and two-loop 
contribution to the selectron mass 
is approximately $m^2_{\tilde{e}^c,0}-k \alpha_1 
\alpha_3 M^2_S$ where $k$ is a numerical factor. 
Substituting the gauge-mediated relation 
$m^2_{\tilde{e}^c,0}\sim \alpha^2_1/
\alpha^2_3 m^2_{\tilde{t}^c,0}$, the combined selectron mass is 
$\alpha^2_1/\alpha^2_3 (m^2_{\tilde{t}^c,0}-k (\alpha_3/ \alpha_1) 
\alpha^2_3 M^2_S)$. Since the combined mass of the stop is
$\sim m^2_{\tilde{t}^c,0}-k^{\prime} \alpha^2_3 M^2_S$, 
the limit for $m^2_{\tilde{t}^c,0}$ 
obtained from the positivity requirement for
$m^2_{\tilde{e}^c}$  
is comparable or larger than the constraint 
obtained from requiring that $m^2_{\tilde{t}^c}$ remains 
positive.
For example, with $n_5=2$, $n_{10}=0$ and $M_S \sim 25$ TeV,
the right-handed slepton constraint 
requires that $m_{\tilde{t}^c,0}\sim$ 1.1 TeV. For $n_{10}$=2, 
$n_5$=0 and $M_S \sim 25$ TeV, 
$\tilde{L}_3$ is driven negative and 
implies that $m_{\tilde{t}^c,0}\sim$ 1 TeV. From 
Figure \ref{m2lgm} we find that these results 
are comparable
to the direct constraint on $m_{\tilde{t}^c,0}$ obtained by
requiring that colour is not broken.

\begin{figure}
\centerline{\epsfxsize=0.8\textwidth \epsfbox{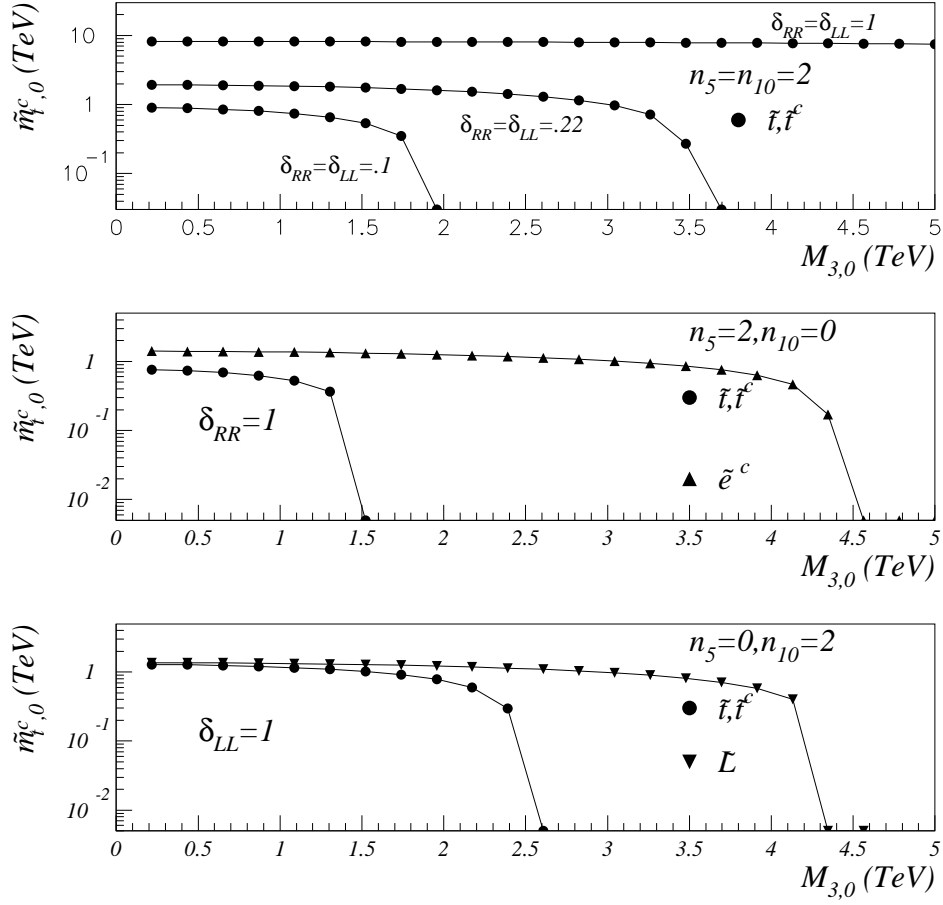}}
\caption{Limits for $m_{\tilde{t}^c,0}$ 
from the requirement that the stop and slepton mass squared are
positive at the weak scale while suppressing
$\Delta m_K$, for different
values of $(n_5,n_{10})$, and $(\delta_{LL},\delta_{RR})$. 
The regions below the curves
are excluded.
Low-energy
gauge-mediated
supersymmetry breaking mass relations between the light scalars 
and $\tan \beta=$2.2 
are assumed.}
\protect\label{m2lgmmk}
\end{figure}

The positivity analysis only constrains 
$m_{\tilde{t}_i,0}/M_S$ for a fixed value 
of $M_{3,0}/M_S$.
To directly limit the initial scalar masses some 
additional information is 
needed. 
This is provided by the measured value of $\Delta m_K$. 
If some mixing 
and degeneracy between the first two generation scalars
is assumed, parameterized by 
$(\delta_{LL},\delta_{RR})$, 
a  
minimum value for $M_S$ is 
obtained by requiring that the supersymmetric contribution 
to $\Delta m_K$ does not exceed the measured value. 
We use the results
given in Section \ref{setup} to calculate this minimum 
value. 
This result together with 
the positivity analysis then 
determines a minimum value for $m_{\tilde{t}^c,0}$ for a 
given initial gluino mass $M_{3,0}$. 
The RG analysis 
is repeated with $\mu_S=M_S$, rather than $\mu_S$=50 TeV. 
We only present the 
results found by assuming GMSB mass relations between the scalars. 
These results are 
shown in Figure \ref{m2lgmmk}. 
The mass limits for other $\tilde{f}_i$ are easily obtained from the
information provided in Figure \ref{m2l} 
and Table \ref{mktable} and are not shown.
From Figure \ref{m2lgmmk} we find that for $(n_5,n_{10})$
$=(2,2)$ and $M_{3,0}$ less than 2 TeV, $m_{\tilde{t}^c,0}$ 
must be larger than $8$ TeV for 
$\sqrt{\delta_{LL} \delta_{RR}}=1$, and larger than $1.8$ TeV for 
$\sqrt{\delta_{LL} \delta_{RR}}=.22$. 
This results in $\Delta(m_Z^2,m_{\tilde{t} ,0}^2)$ of
2000 and 120, respectively.
In this case both the squark and selectron limits for $m_{\tilde{t}^c,0}$ 
are comparable.
The limits for other choices for $\sqrt{\delta_{LL} \delta_{RR}}$ are 
obtained from Figure \ref{m2lgmmk} by a simple scaling, since to a good
approximation $\Delta m_K \sim \delta_{LL} \delta_{RR}/M^2_S$. 
For the cases $(n_5,n_{10})=(2,0)$ and $(0,2)$, the corresponding limits 
are much weaker. In the case $(n_5,n_{10})=(2,0)$, 
for example, only for $\delta_{RR}\sim 1$ does the selectron mass
limit require that $m_{\tilde{t}^c,0} \sim $1 TeV. The limits for a smaller 
value of $\delta$ are not shown.

We conclude with some comments about how these results change if $CP$ violation is present in these theories with $O(1)$ phases.
Recall from Section \ref{setup} that for the same choice of 
input parameters, the limits 
on the initial stop masses increases by about a 
factor of 12. This may be interpreted in one of two ways. 
Firstly, this constrains those models that were relatively 
unconstrained by the $\Delta m_K$ limit. We concentrate 
on those models with $n_5=2$ and $n_{10}=0$, since this case 
is the most weakly constrained by the combined $\Delta m_K$ and 
positivity analysis. The conclusions for other models will be 
qualitatively the same.  We find from Figure \ref{m2lgmmk} the 
limit $m_{\tilde{t}^c,0}>$1 TeV \footnote{For GMSB relations only. 
The direct constraint on the stop masses is slightly weaker.} 
is only true if $\delta_{RR} \sim O(1)$. Smaller values 
of $\delta_{RR}$ do not require large initial stop masses. 
From the $CP$ violation constraint, however, smaller values 
for $\delta_{RR}$ are now constrained. For example,  
if $\delta_{RR}\sim$0.1 and $O(1)$ phases are present, 
then $m_{\tilde{t}^c,0}>$1 TeV is 
required. Secondly, the strong constraint from $\epsilon$ 
could partially or completely compensate a weakened constraint 
from the positivity analysis. This could occur, for example, 
if in an actual model the negative two-loop contribution 
to the stop mass squared for the same initial input parameters 
is smaller than the estimate used here. For example, if the 
estimate of the two-loop contribution in an actual model 
decreases by a factor       
of $\sim (12.5)^2$ and $O(1)$ phases are present, the 
limit in this case from $\epsilon$ for the same 
$\delta$ is identical to the values presented in Figure \ref{m2lgmmk}.

\section{High Scale 
Supersymmetry Breaking}
\label{highsusy}

In this section, we consider the case in which SUSY 
breaking is communicated to the
MSSM fields at a high energy scale, that is taken to 
be \footnote{This choice for the high scale is
done to remain agnostic about any physics appearing between the
Grand Unification scale and the Planck scale. This also results in a 
conservative 
assumption, since the negative two-loop contribution is smaller with 
$M_{SUSY}=M_{GUT}$.} $M_{GUT} = 2 \times
10^{16}$ GeV. 
In this case, the negative contribution 
of the heavy scalar
soft masses to the soft mass squareds of the light 
scalars is enhanced by $\sim \ln (M_{GUT}/\hbox{50 TeV})$,
since the heavy scalar soft masses contribute to the 
RGE from $M_{GUT}$ to
mass of the heavy scalars. It is clear that as the scale of SUSY 
breaking is lowered the
negative contribution of the heavy scalar
soft masses reduces.

This scenario was investigated in Reference \cite{nima}, and we  
briefly discuss the difference between that analysis and the results 
presented here.
In the analysis of Reference \cite{nima}, the authors made the 
conservative choice of neglecting $\lambda_t$ in the 
RG evolution.  The large 
value of $\lambda_t$ can change the analysis, and 
it is included here. 
We find  
that for some pattern of initial stop and up-type Higgs                          scalar masses, e.g. universal scalar
masses, this effect increases the constraint on the stop 
masses by almost 
a factor of two.  
This results in an increase of a factor of 3-4 in the amount of 
fine tuning required to obtain the correct $Z$ mass. 
Further, in combining the positivity 
analysis with the constraints from the 
$\Delta m_K$ analysis, the QCD corrections to the Flavour Changing
Neutral Current (FCNC) operators has been 
included, as discussed in Section \ref{setup}. In the case $(n_5,n_{10}) 
=(2,2)$, this effect alone 
increases the positivity limit by a factor of $\sim 2-3$. 
The combination of these two elements imply that the positivity constraints 
can be quite severe. 

We proceed as follows. First, we solve the RGEs from 
$M_{GUT}$ to $\mu _S$ where
the heavy scalars are decoupled. At this scale, we add 
the finite parts of
the two-loop diagrams. Next, we RG scale  
(without
the heavy scalar terms in the RGEs) from
$\mu _S$ to $\mu _G$ using these new boundary conditions.
Except where stated otherwise, 
the scales $\mu_S$ and $\mu_G$ are fixed to be 50 TeV and 1 TeV, 
respectively.
  
\begin{figure}
\centerline{\epsfxsize=0.5\textwidth \epsfbox{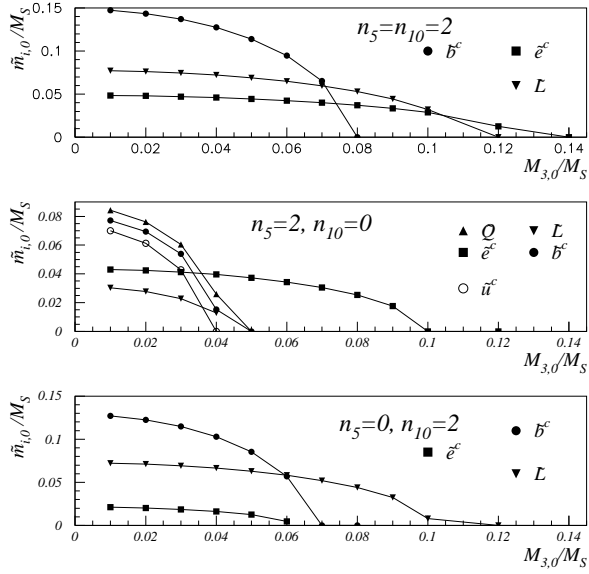}}
\caption{Limits for $m_{\tilde{f}_i,0}$ for different values of
$(n_5,n_{10})$ from the requirement
that the mass squareds are positive at the weak scale,
assuming 
a supersymmetry breaking scale of $M_{GUT}$. The regions below the lines 
are excluded. }
\protect\label{hm2}
\end{figure}

For $\tilde{f} _i \neq \tilde{t}$, $\tilde{t}^c$ we find,
\begin{eqnarray}
m^2_{\tilde{f} _i}(\mu _G) &=&  m^2_{\tilde{f} _{i,0}} +
( 2.84 C^i_3 + 0.639 C_2^i + 0.159 Y^2_i )
M^2_{3,0} +c_D Y_i D_{Y,0} \nonumber \\
 & & -( 4.38 C^i_3 + 1.92 C^i_2 + 0.622 Y^2_i )
\frac{1}{2} (n_5 +3n_{10}) \times 10^{-3} M^2_S
 \nonumber \\
 & & -  0.829 (n_5 - n_{10}) Y_i \times 10^{-3} M^2_S \nonumber \\
& & +(n_5-n_{10})\left( 17.2 M^2_{3,0}+(.226-0.011(n_5+3 n_{10}))M^2_S\right)
Y_i \times 10^{-3}.
\label{highscale}
\end{eqnarray}
These results agree with Reference \cite{nima} for the same 
choice of input parameters. The term proportional to $D_{Y,0}$, and the 
terms in the last line result from 
integrating the one-loop hypercharge $D-$term. In this case $c_D=-0.051$. 
As in the previous Section, the numerical coefficients in 
Equation(\ref{highscale}) depend on
$(n_5,n_{10})$ through the gauge coupling evolution, and the 
numbers in
Equation(\ref{highscale}) are for $(n_5,n_{10}) = (2,0)$ 
\footnote{The numerical results presented in Figure \ref{hm2} 
include this dependence.}. Requiring the soft masses squared to be 
positive constrains $\tilde{m}^2_{i,0}=m^2_{i,0}+c_D Y_i D_{Y,0}$. 
In Figure \ref{hm2} we plot the values of $\tilde{m}_{\tilde{f} _{i,0}}/M_S$ 
that determine 
$\tilde{m}^2_{\tilde{f} _i}(\mu _G) = 0$ as a function of $M_3/M_S$,
for $\tilde{f}_i = \tilde{L}_i$, $\tilde{Q}_i$, 
$\tilde{u}^c_i$, $\tilde{d}^c_i$ 
and $\tilde{e}^c_i$. We emphasize that the results presented in Figure
\ref{hm2} are independent
of any further limits that FCNC or fine tuning considerations may 
imply, and are thus  
useful constraints on any model building attempts.

For the stops, the numerical solutions to the RGEs for $\tan \beta=2.2$ are
\begin{eqnarray}
m^2_{\tilde{t}}(\mu_G) & = & - 0.021 A_t^2 +0.068 A_t M_{3,0} +
 3.52 M^2_{3,0} +c_D \frac{1}{6} D_{Y,0} \nonumber \\
 & & - 0.142 (m^2_{H_u,0}+ m^2_{\tilde{t}^c,0}) +
0.858 m^2_{\tilde{t},0} -
 c_L 0.00613 M^2_S \nonumber \\
m^2_{\tilde{t}^c}(\mu_G) & = & - 0.042 A_t^2 + 0.137 A_t M_{3,0} 
+ 2.33 M^2_{3,0} +c_D \frac{-2}{3} D_{Y,0} 
 \nonumber \\
 & & - 0.283 (m^2_{H_u,0}+ m^2_{\tilde{t},0}) +
0.716 m^2_{\tilde{t}^c,0} -
 c_R 0.00252M^2_S \nonumber \\
m^2_{H_u}(\mu_G) & = & - 0.063 A_t^2 + 0.206 A_t M_{3,0} - 
1.72  M^2_{3,0} +c_D \frac{1}{2} D_{Y,0}
\nonumber \\
 & & - 0.425 (m^2_{\tilde{t},0}+ 
m^2_{\tilde{t}^c,0})+
0.574 m^2_{H_u,0} +
 c_H 0.00193 M^2_S
\label{highstop}
\end{eqnarray}
where $(c_L,c_R,c_H) = (1,1,1)$, $(3.57,4.92,5.15)$,
$(2.7,4.16,4.27)$ for $(n_5,n_{10}) = 
(2,0)$, $(2,2)$ and $(0,2)$, respectively. Also, $c_D=-0.051$.
The  
mixed two-loop contribution to the RG evolution is $\propto (n_5-n_{10})$ 
and is not negligible. 
Thus there
is no simple relation between the $c$'s for different values of 
$n_5$ and $n_{10}$. 
From Equations \ref{muEW} and \ref{ft} we find that to keep $m^2_Z$ fine tuned 
at less than $1\%$ ($\Delta \leq 100$) in each of the bare
 parameters, 
we must have: 
$\mu \ltap$ 460 GeV; $M_{3,0} \ltap 300
$ GeV; $m_{\tilde{t}_i,0} \ltap .87$ TeV; $m_{5,i} \ltap 
16$ TeV; and $m_{10,i} \ltap 10$ TeV, for $(n_5,n_{10})=(2,2)$. The fine tuning 
of the $Z$ mass with respect to the heavy scalars is discussed in 
\cite{dimopoulos}. 
Finally, for other values of these parameters
the fine tuning increases as
$\Delta= 100 \times \tilde{m}^2/ \tilde{m}^2_0$, 
where $\tilde{m}_0$ is the value
of $\tilde{m}$ that gives $\Delta=100$.

As in Section \ref{lowsusy}, we rewrite Equations \ref{highstop} in 
terms of $\tilde{m}^2_{i,0}=m^2_{i,0}+c_D Y_i D_{Y,0}$. This is 
equivalent to setting $D_{Y,0}=0$ in Equations \ref{highstop}, and 
relabeling $m^2_{i,0} \rightarrow \tilde{m}^2_{i,0}$. In what follows, 
we use the positivity analysis to constrain $\tilde{m}^2_{i,0}$. 
Since $c_D$ is small and $D_{Y,0} \sim O(m^2)$, the difference between 
$\tilde{m}^2_{i,0}$ and $m^2_{i,0}$ is small. To simplify the 
notation, in the remainder of this Section 
we will also remove the tilde from $\tilde{m}^2_{i,0}$.

\begin{figure}
\centerline{\epsfxsize=0.6\textwidth \epsfbox{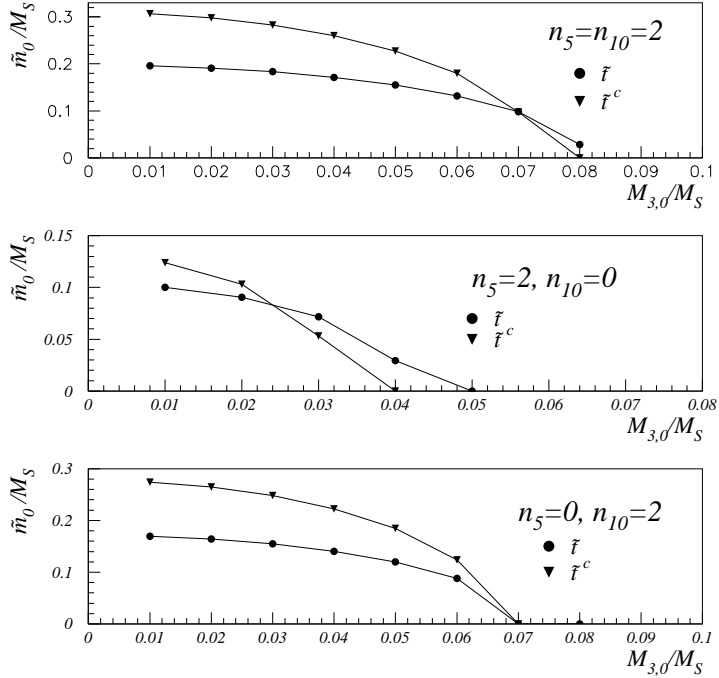}}
\caption{Limits for $\tilde{m}_0/M_S$ 
from the requirement
that the stop mass squareds are positive at the weak scale,
for 
$\tan \beta=2.2$, $A_{t,0}=0$ and assuming universal scalar masses 
at $M_{GUT}$ for the stop and Higgs scalars. 
The regions below the curves are
excluded.}
\protect\label{m0}
\end{figure}

\begin{figure}
\centerline{\epsfxsize=0.6\textwidth \epsfbox{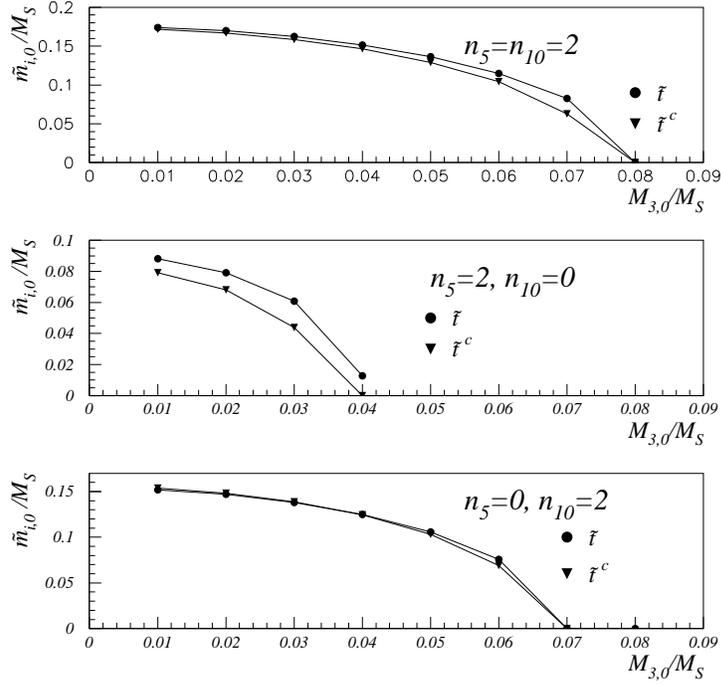}}
\caption{Limits for $m_{\tilde{t},0}/M_S$, $m_{\tilde{t}^c,0}/M_S$,
from the requirement
that the stop mass squareds are positive at the weak scale, 
for $M_{SUSY}=M_{GUT}$, $\tan \beta=2.2$ and assuming that 
$m^2_{H_u,0}=0.$ The value of $A_{t,0}$ is chosen to maximize the 
value of the stop soft masses at the weak scale. 
The regions below the curves are
excluded.}  
\protect\label{mhu0}
\end{figure}

\begin{figure}
\centerline{\epsfxsize=0.6\textwidth \epsfbox{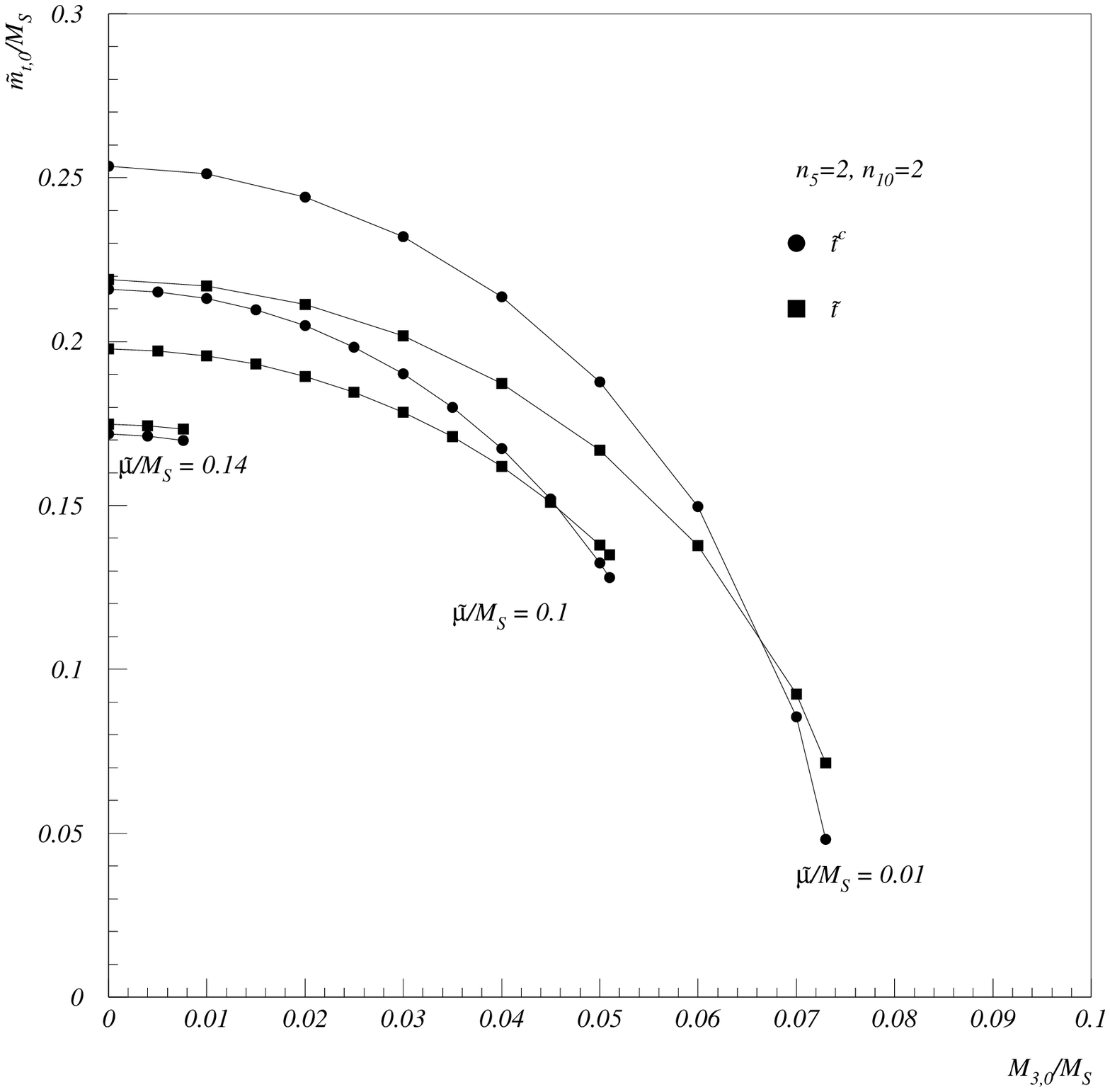}}
\caption{Limits for $m_{\tilde{t},0}/M_S$, $m_{\tilde{t}^c,0}/M_S$,
from the requirement
that the stop mass squareds are positive at the weak scale,
for $(n_5,n_{10})=(2,2)$, 
$M_{SUSY}=M_{GUT}$, $\tan \beta=2.2$, and different 
values of $\tilde{\mu}/M_S$. The contours end at that value of 
$M_{3,0}/M_S$ that gives $m_{H_u,0}/M_S=0$.
The value of $A_{t,0}$ is chosen to maximize the
value of the stop soft masses at the weak scale. 
The regions below the lines are excluded.}
\protect\label{mu}
\end{figure}

As was also discussed in Section \ref{lowsusy}, some relations 
between $m^2_{\tilde{t},0}$,
$m^2_{\tilde{t}^c,0}$ and $m^2_{H_u,0}$ are needed to obtain 
a constraint from
Equation(\ref{highstop}), using 
$m^2_{\tilde{t}}(\mu _G) >0$ and
$m^2_{\tilde{t}^c}(\mu _G) >0$. 
We discuss both model-dependent and model-independent constraints on 
the initial values of the stop masses. 
The outline of the rest of this Section 
is as follows. First, we assume universal boundary conditions. These 
results are presented in Figure \ref{m0}. 
Model-independent constraints are obtained
by the following.
We assume that 
$m^2_{H_u,0}=0$
and choose $A_{t,0}$ to maximize the value of the stop masses at the weak 
scale. These results are presented in Figure \ref{mhu0}. We further 
argue that these constraints represent minimum constraints as long 
as $m^2_{H_u,0} \geq 0$. 
To obtain another set of model independent constraints, we use  
the electroweak symmetry breaking
relation to eliminate $m^2_{H_u,0}$ in favour of $\mu$.
Then we present the positivity limits
for different values of $\tilde{\mu}/M_S$, where $\tilde{\mu}^2=\mu^2+
\frac{1}{2}m^2_Z$, and 
assume that $m^2_{H_d,0}=0$ to 
minimize the value of $\mu$ \footnote{Strictly speaking, this last 
assumption is unnecessary. Only the combination 
$\tilde{\mu}^2_H \equiv 
\tilde{\mu}^2-m^2_{H_d,0}/ \tan ^2 \beta$ appears in our analysis. 
Thus for $m^2_{H_d,0} \neq 0$ our results are unchanged if the 
replacement $\tilde{\mu} \rightarrow \tilde{\mu}_H$ is made.}.
These limits are model-independent and are presented in 
Figure \ref{mu},
for the case $n_5 = n_{10} =2$. 
We then combine these analyses with the limits on $M_S$ obtained 
from $\Delta m_K$. We conclude with some discussion about the anomalous 
$D-$term solutions to the flavour problem.

 We first consider universal boundary
conditions for the stop and Higgs masses. 
That is, we assume that $m^2_{\tilde{t},0} 
= m^2_{\tilde{t}^c,0}=
m^2_{H_u,0}= \tilde{m}^2_0$. In Figure \ref{m0} we plot for 
$\tan \beta=2.2$ the 
minimum value of $\tilde{m}_0/M_S$ required to
maintain $m^2_{\tilde{t}}(\mu _G) >0$ and
$m^2_{\tilde{t}^c}(\mu _G) >0$. This value of $\tan \beta$ 
corresponds to $\lambda_t(M_{GUT})=.88$, in the case that $(n_5,n_{10})=(2,0)$. 
For comparison, the results 
gotten assuming $\lambda_t=0$ may be found in Reference \cite{nima}. 
For 
$n_5=n_{10}=2$ we note from Figure \ref{m0} that if $M_S=$ 20 TeV 
and the gaugino masses are small, the limit on the stop mass is 
$m_{\tilde{t}^c,0} \geq$ 6.2 TeV. 
This limit is weakened to 6 TeV if $M_{3,0}\ltap$ 300 GeV 
is allowed. 
Even in this case, 
this large initial stop mass requires a fine tuning  
that in this case is $\Delta\sim (\hbox{6 TeV})^2/m^2_Z \sim$ 4200, i.e.
a fine tuning of $\ltap 10^{-3}$ is needed to obtain the correct $Z$ mass.     
 
We now assume $m^2_{H_u,0}=0$ and choose the initial 
value of $A_{t,0}$ to {\em maximize} the value of 
$m^2_{\tilde{t}_i}(\mu_G)$.
The values  
of $m^2_{\tilde{t},0}$ and $m^2_{\tilde{t}^c,0}$ are 
chosen such that  
$m^2_{\tilde{t}}(\mu _G) >0$ and
$m^2_{\tilde{t}^c}(\mu _G)>0$. We note that in this case 
the constraint is 
weaker because the $\lambda_t$ contribution to the RG 
evolution of the stop masses is less negative. 
These results are plotted in Figure \ref{mhu0}. 

We discuss this case in some more detail and argue that 
the minimum value of $m_{\tilde{t}_i,0}$ obtained in this way will 
be valid for all $m^2_{H_u} \geq0$ and all
$A_{t,0}$. 
Eliminate the $A_{t,0}$ term by choosing $A_{t,0}=
k M_{3,0}$ such that the $A_t$ contributions to $m^2_{\tilde{t}_i}(\mu_G)$ 
is maximized. Other choices for $A_{t,0}$ require larger values for 
$m^2_{\tilde{t}_i,0}$ to maintain $m^2_{\tilde{t}_i}(\mu_G)=0$. 
The value of $k$ is determined by the following. A general 
expression for the value of the soft masses of the stops at the weak scale 
is 
\begin{equation}
 m^2_{\tilde{t}}(\mu_G)=-a A^2_{t,0}+b A_{t,0} M_{3,0}+c M^2_{3,0}+\cdots,
\end{equation}
\begin{equation}
 m^2_{\tilde{t}^c}(\mu_G)=-2a A^2_{t,0}+2b A_{t,0} M_{3,0}+d M^2_{3,0}+\cdots,
\end{equation}       
with $a$, $c$ and $d$ positive. The maximum value of 
$m^2_{\tilde{t}_i}(\mu_G)$ 
is obtained by choosing $A_{t,0}=bM_{3,0}/2a$.  
The value of the stops masses at this 
choice of $A_{t,0}$ are  
\begin{equation}
 m^2_{\tilde{t}}(\mu_G)=(c+\frac{b^2}{4 a}) M^2_{3,0}+\cdots,
\end{equation}
\begin{equation}                                                           
m^2_{\tilde{t}^c}(\mu_G)=(d+ 2 \frac{b^2}{4 a}) M^2_{3,0}+\cdots.
\end{equation}
An inspection of Equation \ref{highstop} gives 
$b=0.068$ and $a=0.021$ for 
$\tan \beta =2.2$. In this case the `best' value for $A_{t,0}$ is 
$A^B_{t,0}\sim 1.6 M_{3,0}$. It then follows 
that the quantity $b^2/4a=0.055$ is a small correction 
to the coefficient of the gaugino contribution in Equation \ref{highstop}.
Thus the difference between the minimum initial stop masses 
for $A_{t,0}=0$ and $A_{t,0}$= 
$A^B_{t,0}$ is small.
Next assume that $m^2_{H_u,0}=0$.
Requiring that both 
$m^2_{\tilde{t}}(\mu _G)=0$ and $m^2_{\tilde{t}^c}(\mu _G)=0$ 
determines a minimum value for $m^2_{\tilde{t},0}$ and 
$m^2_{\tilde{t}^c,0}$.   
Now since the $m^2_{H_u,0}$ contribution to both the
stop soft masses is negative $($see Equation \ref{highstop}$)$, the
minimum values for $m^2_{\tilde{t}_i,0}$ 
found by the preceeding procedure are also minimum values
if we now allow any $m^2_{H_u,0} >0$. 

We conclude that for all $A_{t,0}$ and all $m^2_{H_u,0} \geq0$, 
the limits presented in Figure \ref{mhu0} represent 
lower limits on the initial stop masses if we require that the soft
masses remain positive at the weak scale. Further, the limits in this 
case are quite 
strong. For example, from Figure \ref{mhu0} we find that if 
$M_S \sim$ 20 TeV and $M_{3,0} \sim $ 200 GeV 
(so that $M_{3,0}/M_S \sim$10$^{-2})$,
then the initial stop masses must be greater than 3.5 TeV
in the case that $(n_5,n_{10})=(2,2)$
The results are stronger in a more realistic 
scenario, $i.e.$ $m^2_{H_u,0}>0$.  If, for example, $m^2_{H_u,0} \sim 
m^2_{\tilde{t}^c,0}/9$ the constraints are larger by only a few percent. 
In the case that $m^2_{H_u,0}=m^2_{\tilde{t}^c,0}=m^2_{\tilde{t},0}$, 
presented in Figure \ref{m0}, however, 
the constraint on the initial $\tilde{t}^c$ mass
increases by almost a factor of two. 

To obtain constraints on the initial stop masses we have thus 
far had to assume some relation between $m^2_{H_u,0}$ and 
$m^2_{\tilde{t}^c,0}$; e.g., $m^2_{H_u,0}=0$ or 
$m^2_{H_u,0}=m^2_{\tilde{t}^c,0}$. Perhaps a better approach is 
to use the EWSB relation, Equation (\ref{muEW}), to
eliminate $m^2_{H_u,0}$ in favour of $\mu ^2$. This has the advantage of
being model-independent.   
It is also a useful reorganization of independent parameters since 
the amount of fine tuning required to obtain the correct
$Z$ mass increases as $\mu$ is increased. 
To obtain some limits we choose $m_{H_d ,0}^2 = 0$ 
\footnote{This assumption is unnecessary. 
See the previous footnote.} to minimize the value of $\mu^2$, 
and require that $m^2_{H_u,0}$ is positive. The minimum value of 
$m_{\tilde{t}^c,0}/M_S$ and $m_{\tilde{t},0}/M_S$ 
for different choices of $\tilde{\mu}/M_S$ are gotten by 
solving $m^2_{\tilde{t}^c}(\mu _G) =0$ and 
$m^2_{\tilde{t}}(\mu _G) =0$. These results are presented in 
Figure \ref{mu}. 
In this Figure the positivity constraints terminate at that
value of $M_{3,0}$ which gives
$m^2_{H_u,0}=0$. 

As discussed in the above, reducing the value of $m^2_{H_u,0}$ 
decreases the positivity limit on $m_{\tilde{t}_i,0}$. Consequently  
the fine tuning of $m^2_Z$ with respect to $m^2_{\tilde{t}_i,0}$ is also 
reduced. But using Equations \ref{highstop} and \ref{muEW},
it can be seen that
decreasing $m_{H_u ,0}^2$ while
keeping $m_{\tilde{t}^c}^2(\mu _G) =0$ and
$m_{\tilde{t}}^2(\mu _G) =0$ 
results in a larger $\mu$, thus increasing
the fine tuning
with respect to
$\mu$. 
This can also be seen from Figure \ref{mu}. We find, for example, that if  
$M_{3,0}/M_S \sim 0.01$, 
the small value $\tilde{\mu}/M_S=0.01$ requires 
$m_{\tilde{t}_i,0}/M_S \sim .25$. For $M_S=10$ TeV, this corresponds to
$\mu \sim$ 100 GeV and $m_{\tilde{t}_i,0}\geq$ 2.5 TeV. 
A further
inspection of Figure \ref{mu} shows that for the same 
value of $M_{3,0}/M_S$, 
a value of 
$m_{\tilde{t},0}/M_S =0.17$
is allowed (gotten by decreasing $m^2_{H_u,0}$) 
only if $\tilde{\mu}/M_S$ is increased to .14.  
This corresponds to $\mu=1.4$ TeV for
$M_S=10$ TeV; this implies that $\Delta(m^2_Z; \mu) \sim 930$. 
We find that the limit on the initial
stop masses can only be decreased at the expense of increasing $\mu$.

Finally, the limits become weaker if
$m^2_{H_u,0}<0$. This possibility is theoretically unattractive on two
accounts.
Firstly, a nice feature of supersymmetric extensions to the SM
is that the dynamics of the model, through the
presence of the large top quark Yukawa coupling, naturally leads
to
the breaking of the electroweak symmetry\cite{ross}. This is lost if
electroweak symmetry breaking is already present at the
tree-level.
Secondly, the fine tuning required to obtain the correct $Z$ mass is increased.
From Figure \ref{mu} we infer that while reducing $m^2_{H_u,0}$ below zero
 does 
reduce the limit on the initial stop masses, the value of $\mu$ increases 
beyond the values quoted in the previous paragraph, thus 
further increasing the 
fine tuning of the $Z$ mass. 
This scenario is not discussed any further.

\begin{figure}
\centerline{\epsfxsize=.6\textwidth \epsfbox{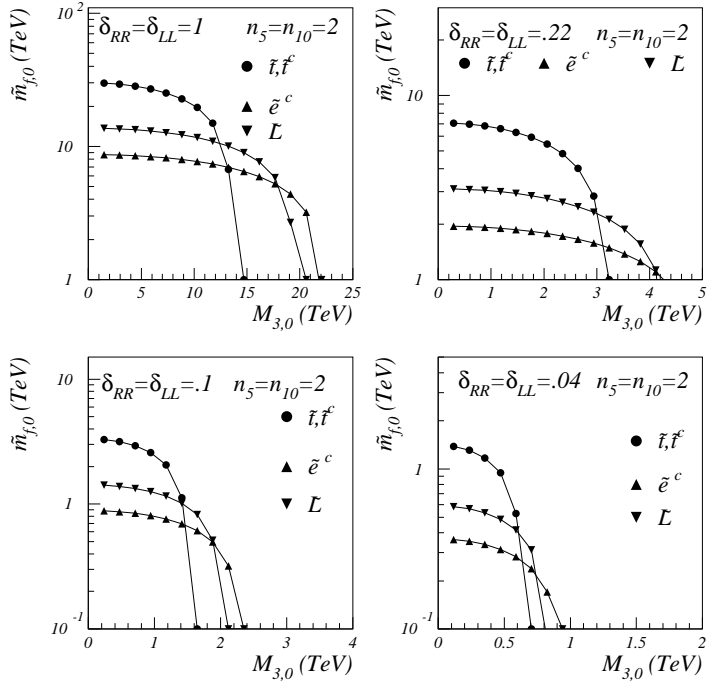}}
\caption{
Limits for $m_{\tilde{t},0}$ and $m_{\tilde{t}^c,0}$,
$m_{\tilde{e}^c}$, and $m_{\tilde{L}}$
from the requirement
that the mass squareds are positive at the weak scale
while suppressing $\Delta m_K$.
It was assumed that $M_{SUSY}=M_{GUT}$, $\tan \beta=2.2$ and that
$m^2_{H_u,0}=0.$ The value of $A_{t,0}$ was chosen to maximize the
value of the stop soft masses at the weak scale. The heavy scalars were 
decoupled at the minimum value allowed by $\Delta m_K$. 
The regions below the lines are excluded.}
\protect\label{mhu0mk}
\end{figure}

\begin{figure}
\centerline{\epsfxsize=.6\textwidth \epsfbox{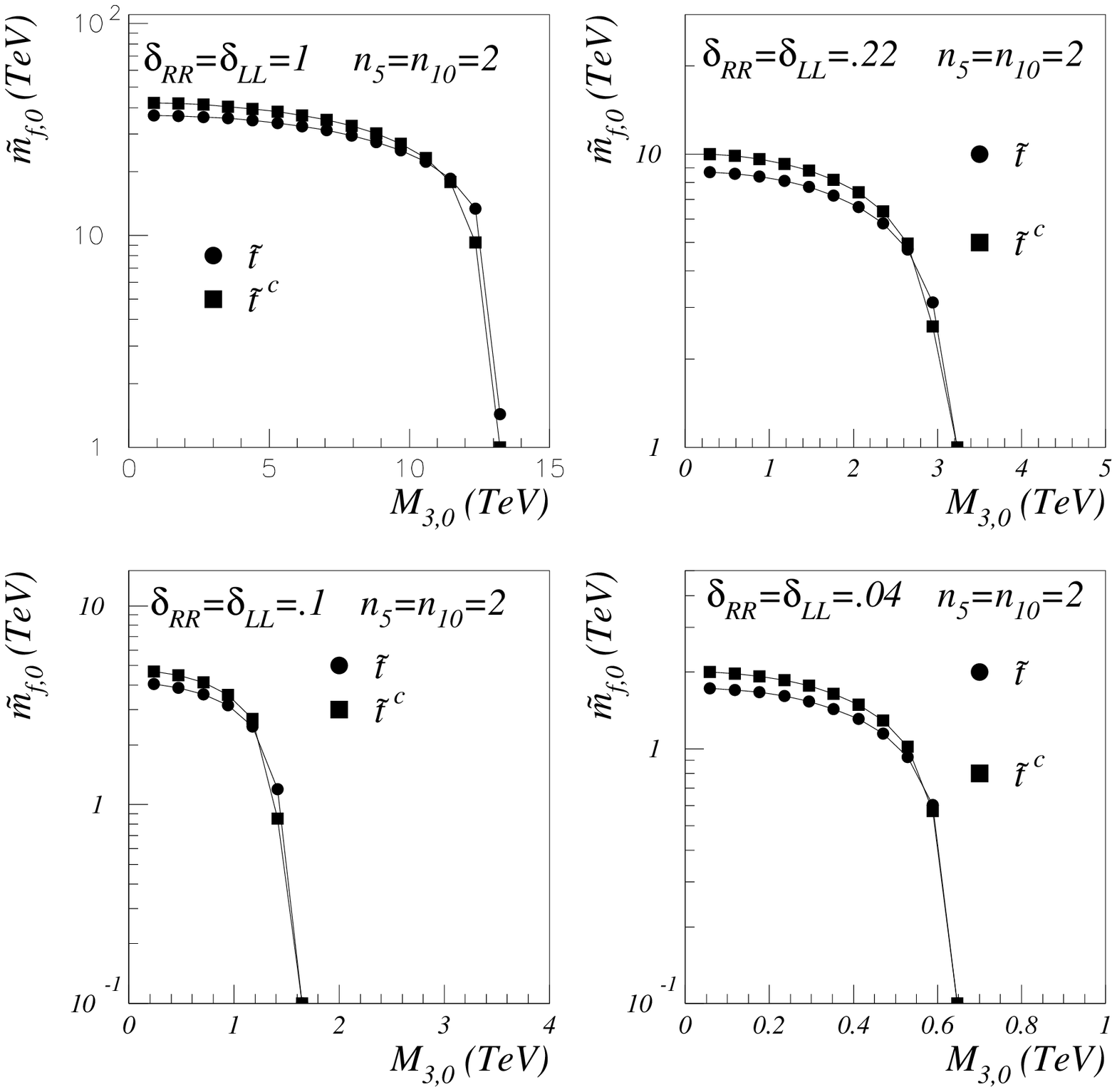}}
\caption{Limits for $m_{\tilde{t},0}$ and $m_{\tilde{t}^c,0}$
from the requirement
that the stop mass squareds are positive at the weak scale
while suppressing $\Delta m_K$.
It was assumed that $M_{SUSY}=M_{GUT}$, $\tan \beta=2.2$ and that
$m^2_{H_u,0}=m^2_{\tilde{t}^c,0}.$ 
The value of $A_{t,0}$ was chosen to maximize the
value of the stop soft masses at the weak scale. The heavy scalars were
decoupled at the minimum value allowed by $\Delta m_K$. 
The regions below the lines are excluded.}
\protect\label{mhu3mk}
\end{figure}

\begin{figure}
\centerline{\epsfxsize=.6\textwidth \epsfbox{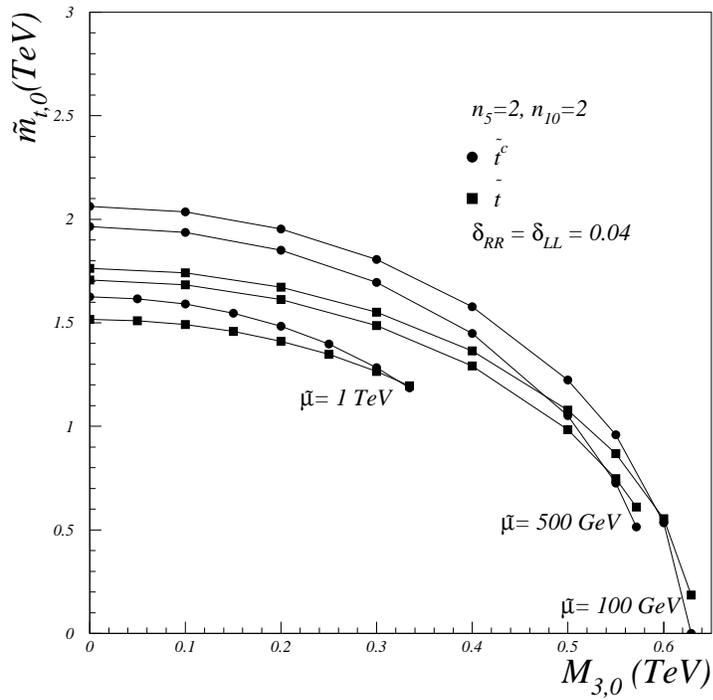}}
\caption{Limits for $m_{\tilde{t},0}$ and $m_{\tilde{t}^c,0}$
from the requirement
that the stop mass squareds are positive at the weak scale
while suppressing $\Delta m_K$,
for $(n_5,n_{10})=(2,2)$, $\sqrt{\delta_{LL} \delta_{RR}}=0.04$,
and different values of $\mu$.
The contours terminate at $m^2_{H_u,0}=0$.
It was assumed that $M_{SUSY}=M_{GUT}$ and $\tan \beta=2.2$. 
The value of $A_{t,0}$ was chosen to maximize the
value of the stop soft masses at the weak scale. The heavy scalars were
decoupled at the minimum value allowed by $\Delta m_K$. 
The regions below the lines are excluded.}
\protect\label{mu2}
\end{figure}

\begin{figure}
\centerline{\epsfxsize=.6\textwidth \epsfbox{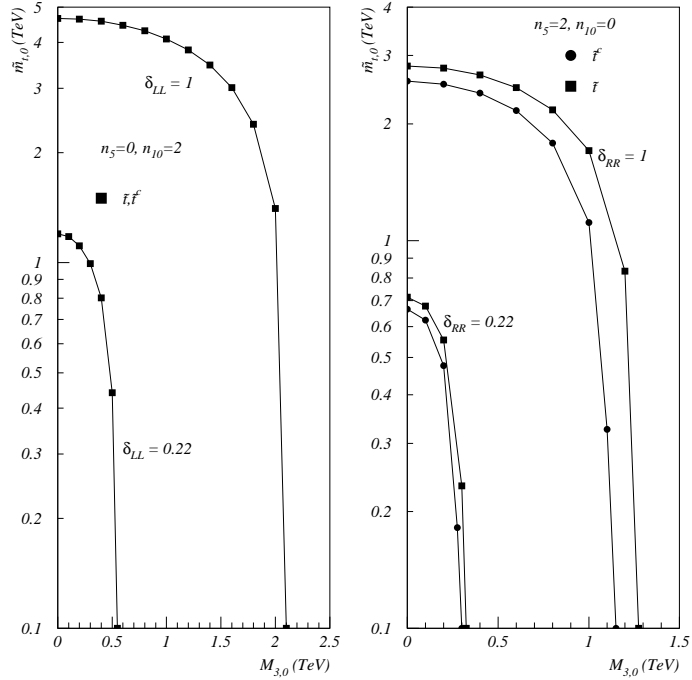}}
\caption{Limits for $m_{\tilde{t},0}$, $m_{\tilde{t}^c,0}$
from the requirement
that the stop mass squareds are positive at the weak scale
while suppressing $\Delta m_K$,
for the cases $(n_5,n_{10})=(2,0)$ and $(0,2)$.
It was assumed that $M_{SUSY}=M_{GUT}$, $\tan \beta=2.2$ and that
$m^2_{H_u,0}=0.$ The value of $A_{t,0}$ was chosen to maximize the
value of the stop soft masses at the weak scale. The heavy scalars were
decoupled at the minimum value allowed by $\Delta m_K$. 
The regions below the lines are excluded.}
\protect\label{mhu0mk2}
\end{figure}

We now combine the positivity analysis of this Section with the results 
of Section \ref{setup} to place lower limits on the soft 
scalar masses.
For given values of $\delta_{LL}, \delta_{RR}$, 
a minimum value of $M_S$, $M_{S,min}$, is 
found using the results of Section \ref{setup}. 
This is combined with the 
positivity analysis in Figure \ref{mhu0}, to produce 
the results shown in Figure \ref{mhu0mk}. We also show other 
limits gotten by assuming $m^2_{H_u,0}=m^2_{\tilde{t}^c,0}$. 
These results are presented in Figure \ref{mhu3mk}.
In Figure \ref{mu2} we also present the stop mass limits for 
different values of $\mu$, and restrict to 
$m^2_{H_u,0} \geq0$ and $\sqrt{\delta_{LL} \delta_{RR}}=0.04$. 
In all cases the heavy scalars 
were decoupled at $M_{S,min}$, rather than 
50 TeV, and so the positivity analysis was 
repeated. The value of $A_{t,0}$ was chosen 
to maximize the value of the stop masses at the weak scale. 
For completeness, the results for the 
cases $(n_5,n_{10})=(2,0)$ and 
$(0,2)$ and $m^2_{H_u,0}=0$ are presented in Figure \ref{mhu0mk2}.  
We repeat that the minimum allowable values for the 
stop masses consistent with 
$m^2_{H_u,0}>0$, gotten by setting $m^2_{H_u,0}=0$, are given in 
Figures \ref{mhu0mk} and \ref{mhu0mk2}.

We next briefly discuss some consequences of this numerical analysis. 
We concentrate on the case $n_5=n_{10}=2$, since this is the relevant
case to consider if the supersymmetric flavour problem is 
explained by decoupling
the heavy scalars.
Other choices for $n_5$ and $n_{10}$ requires
additional physics to explain the required degeneracy or
alignment of any
light non-third generation scalars. 
From Figures \ref{mhu0mk} and \ref{mhu3mk} we find that for 
$\sqrt{\delta_{LL} \delta_{RR}}=0.22$ and $M_{3,0} \leq$ 1 TeV, 
$m_{\tilde{t}_i,0} \ltap$7 TeV is required.  
If instead we 
restrict both $\Delta(m^2_Z;M^2_S)$ and $\Delta(m^2_Z; M_{3,0})$ 
to be less than 100,  
then we must have
$M_S \ltap$ 10 TeV and $M_{3,0} \ltap$ 300 GeV. 
To not be excluded by 
$\Delta m_K$, we further require 
that $\sqrt{\delta_{LL} \delta_{RR}}\ltap .06$.  
For this value of $\sqrt{\delta_{LL} \delta_{RR}}=0.06$, a minimum 
value for $m_{\tilde{t},0}$ of $\sim$1.5$-$2.5 TeV is gotten by
rescaling the results in Figures \ref{mhu0mk} 
and \ref{mhu3mk} for $\sqrt{\delta_{LL} \delta_{RR}}=0.04$ by an amount 
$0.06/0.04$.  
The range depends on the 
value of $m^2_{H_u,0}$, with the lower (upper) limit 
corresponding to $m^2_{H_u,0}=0$ $(m^2_{\tilde{t}^c,0})$. 
Thus $\Delta(m^2_Z; m^2_{\tilde{t}_i,0})\sim 400-800$. This fine tuning 
can be reduced only by either increasing $M_{3,0}-$which 
increases $\Delta(m^2_Z,M_{3,0})$ beyond 100$-$or by 
reducing $M_S-$ which requires 
a smaller value for $\sqrt{\delta_{LL} \delta_{RR}}$. 
We conclude that unless $\sqrt{\delta_{RR} \delta_{LL}}$ is 
naturally small, decoupling the 
heavy scalars does not provide a natural
solution to the flavour problem. 

To conclude this Section we 
discuss the constraint this analysis implies for those models 
which generate
a split mass spectrum between different generations
through the $D$-term contributions of the anomalous
$U(1)$
gauge symmetry\cite{dvali,mohapatra,nelson}.
In the model of set D of \cite{nelson},
there are two ${\bf \bar{5}}$s at $7$ TeV and $6.1$ TeV and
two ${\bf 10}$s at $6.1$ and $4.9$ TeV, respectively, so that
$\Delta m_K$ is suppressed. These values must be increased 
by a factor of 2.5 
to correct for the QCD enhancement of the SUSY 
contribution to $\Delta m_K$, as discussed in Section
\ref{setup}.   
To obtain a conservative 
bound on the initial stop masses from the positivity requirement, 
we first assume that all the 
heavy scalars have a common mass $M_S=2.5 \times 5$TeV$=12.5$ TeV.
(It would have been 5 TeV without the QCD correction.)
Then assuming a weak scale value of the gluino mass that is less than 
$710$ GeV and 
setting $m^2_{H_u,0}=0$ $(m^2_{\tilde{t}^c,0})$, we find from 
Figure \ref{mhu0} $($\ref{m0}$)$ that $m_{\tilde{t} ,0}\geq 
2.1$ $(3.6)$ TeV is required.  
This leads to $\Delta(m^2_Z; m^2_{\tilde{t} ,0}) \geq
580$ $ (1700)$.
To obtain a better bound, we repeat our 
analysis using $n_5 m^2_5+\hbox{3} n_{10} m^2_{10}=((\hbox{7 TeV})^2+ 
(\hbox{6.1 TeV})^2+\hbox{3}\times (\hbox{6.1 TeV})^2+
\hbox{3}\times (\hbox{4.9 TeV})^2)\times (\hbox{2.5})^2$. It is possible to 
do this since only this 
combination appears in the RG analysis for $(n_5,n_{10})=(2,2)$. 
We find (assuming
$m_{H_u ,0}^2 =0$ and the gluino mass at the weak scale
is less than $710$ GeV) that $m_{\tilde{t} ,0}
\stackrel{>} \sim 2.4$ TeV.
In the model of \cite{mohapatra},
$\delta_{RR} \approx \delta_{LL} \approx 0.01$. To obtain a limit on the 
initial stop masses, we use the bound obtained from either 
Figures \ref{mhu0mk} or \ref{mhu3mk} for 
$\delta_{RR}=\delta_{LL} \approx 0.04$, 
and divide the limit by a factor of 4.
By inspecting these Figures we find that this model is only weakly 
constrained, even if $ m^2_{H_u,0} \sim m^2_{\tilde{t},0}$. We now 
discuss the limits in this model  
when $O(1)$ $CP$ violating phases are present. 
To obtain the minimum value of 
$M_S$ in this case, 
we should multiply the minimum value of $M_S$ obtained from 
the $\Delta m_K$ constraint 
for $\delta_{LL}=\delta_{RR}=0.04$ by 12.5/4; dividing 
by 4 gives the result for $\delta_{LL}=\delta_{RR}=0.01$, and 
multiplying by 12.5 gives the constraint on $M_S$ from  
$\epsilon$. 
The result is $M_S\gtap$ 23 TeV. 
Next, we assume that $M_{3,0}$ is less than 300 GeV, so that the 
value of the gluino mass at the weak scale is less than 710 GeV. 
This gives $M_{3,0}/M_S \leq0.013$. Using these values of $M_{3,0}$ and 
$M_S$, an inspection of Figures \ref{m0} and \ref{mhu0} implies that 
$m_{\tilde{t},0}$ must be larger 
than 3.9 TeV to 6.9 TeV, depending on the value 
of $m^2_{H_u,0}$. This gives 
$\Delta(m^2_Z; m^2_{\tilde{t} ,0}) \geq 2000$.
In the model of \cite{dvali},
$M_{3,0}/M_S \approx 0.01$ and $m_{\tilde{f} ,0}/M_S
\approx 0.1$.
Inspecting Figures \ref{m0} and \ref{mhu0} we find that these values are
excluded for $(n_5,n_{10})=(2,2)$ and $(0,2)$. The case $(2,0)$ is 
marginally allowed.
The model of \cite{dvali} with 
$(n_5,n_{10})=(2,2)$ and $\lambda_t=0$ was also excluded 
by the analysis of Reference \cite{nima}.  

\section{\bf Using Finetuning to Constrain $\delta$}
\label{sectionft}

In this section,
we vary the messenger scale, $M_{SUSY}$, between the GUT scale
and a low scale $\sim$ 50 TeV, and restrict the boundary values
of the stop and gluino masses so that EWSB is not fine tuned.
This gives us an upper limit to $\delta$ if we require both
positivity of the stop mass squareds at the weak scale and
suppression of $\Delta m_K$. In other words,
we determine the values for $(\delta, M_{SUSY})$  
which are allowed by the following 
requirements: 1. Suppression of the SUSY contribution to
$\Delta m_K$ by making the mass of the first two generation scalars, $M_S$,
large. 2. Positivity of the stop mass squareds and 3. Fine tuning in
electroweak
symmetry breaking does not exceed 1$\%$ or 10$\%$ ({\it i.e.,}
both $\Delta(m^2_Z,m^2_{\tilde{t},0})$ and $\Delta(m^2_Z,M_{3,0})$ are smaller
than either 100 or 10).

An upper limit to $\delta$ satisfying the above requirements
is obtained as follows.
For a given $M_{SUSY}$ we compute, using Equations \ref{ft} and \ref{muEW}, the
boundary values of the stop mass, $m_{\tilde{t},max}$, and the gluino mass,
$M_{3,max}$, such that both $\Delta(m^2_Z,m^2_{\tilde{t},0})$ and
$\Delta(m^2_Z,M_{3,0})$ are equal to some maximum value
$\Delta_{max}$ which is chosen to be 100 or 10.
\footnote{In computing the
$\Delta$'s, $\tan \beta$, in addition to $m_{H_u}^2(m_Z)$,
should be regarded as a function of the bare
parameters. However, this additional contribution
to the $\Delta$'s is small for $\tan \beta \stackrel{>}{\sim} 2$
and also makes the magnitude of $\Delta$ larger. We neglect this
dependence which is a conservative choice.}
Substituting these values of the bare stop\footnote{Strictly speaking, 
we should translate  
the upper bound on $m^2_{\tilde{t}_i,0}$ into an upper bound 
on $\tilde{m}^2_{\tilde{t}_i,0}$ using 
$\tilde{m}^2_{\tilde{t}_i,0}=m^2_{\tilde{t}_i,0}+c_D Y_{\tilde{t}_i} D_{Y,0}+
Y_{\tilde{t}_i} \zeta_D$, {\it i.e.}, 
to that combination appearing in the positivity constraint. Instead, we 
use the same bound for both $m^2_{\tilde{t}_i,0}$ and 
$\tilde{m}^2_{\tilde{t}_i,0}$. This is reasonable, 
since $c_D$ is generally small $(\ltap 0.05)$, and $D_{Y,0} \sim O(m^2)$. In 
any case, this effect is in the opposite direction for 
$\tilde{t}$ and $\tilde{t}^c$. In the 
case that $\zeta _D\neq 0$, a slightly larger $(O(30 \%))$ value 
for $\delta$ may be 
allowed as compared to $\zeta_D=0$. This is because if $\zeta_D <0$, 
the maximum value for $\tilde{m}^2_{\tilde{t}^c,0}$ is larger than 
$ m^2_{\tilde{t},max}$. 
This, in turn, allows for a larger value of $M_S$, and hence $\delta$. 
Naturalness considerations limit $| \zeta_D|$, though. 
The EWSB relation for $m^2_Z$, Equation 
\ref{muEW}, contains a term linear in $\zeta_D$. Requiring that 
$\Delta(m^2_Z, \zeta_D) <100$ implies that 
$| \zeta_D| \ltap \zeta_{D,max} \equiv (900$ GeV$)^2$. 
Thus for a high scale of supersymmetry breaking, 
the upper bound on $\tilde{m}^2_{\tilde{t}^c,0}$ may be increased to  
$\tilde{m}^2_{\tilde{t}^c,0}\sim m^2_{\tilde{t},max}
+\frac{2}{3} \zeta_{D,max} \sim \frac{5}{3} m^2_{\tilde{t},max}$, while 
maintaining $\Delta(m^2_Z,m^2_{\tilde{t}^c,0})=\Delta(m^2_Z, \zeta_D)=100$. 
This roughly translates into 
an increase of $\sim \sqrt{5/3}=1.3$ in the limit to $\delta$. The actual 
limit will be smaller, since with this choice of sign for $\zeta_D$, the 
positivity constraint for the left-handed stop is now stronger. 
It is thus reasonable to require 
that the maximum value of $\tilde{m}^2_{\tilde{t}_i,0}$
be comparable to $m^2_{\tilde{t}_i,max}$.} 
and gluino masses
into the expression for the weak-scale value of the stop mass squared, 
 we determine the maximum value of $M_S$, $M_{S,max}$,
 such that
the stop mass squareds at the weak scale are positive. Using this value for 
$M_S$ and the analysis described in Section \ref{mk}, an upper value to 
$\delta$ is gotten from the $\Delta m_K$ constraint. 
This value of $\delta$ and $M_{SUSY}$
then satisfies the
above-mentioned three requirements. This can be seen as follows. 
For the given $M_{SUSY}$,
if $\delta$ is larger than this limit, then to suppress $\Delta m_K$,
$M_S$ has to be larger than $M_{S,max}$. But, then to keep the stop
mass squareds positive at the weak scale,
the boundary value of either the stop or the gluino
mass has to increase beyond $m_{\tilde{t},max}$ or $M_{3,max}$
respectively, leading to $\Delta(m^2_Z,m^2_{\tilde{t},0})$ or
$\Delta(m^2_Z,M_{3,0})$ larger than $\Delta _{max}$, {\it i.e.}, increasing
the fine tuning in EWSB.

\begin{figure}
\centerline{\epsfxsize=0.8\textwidth \epsfbox{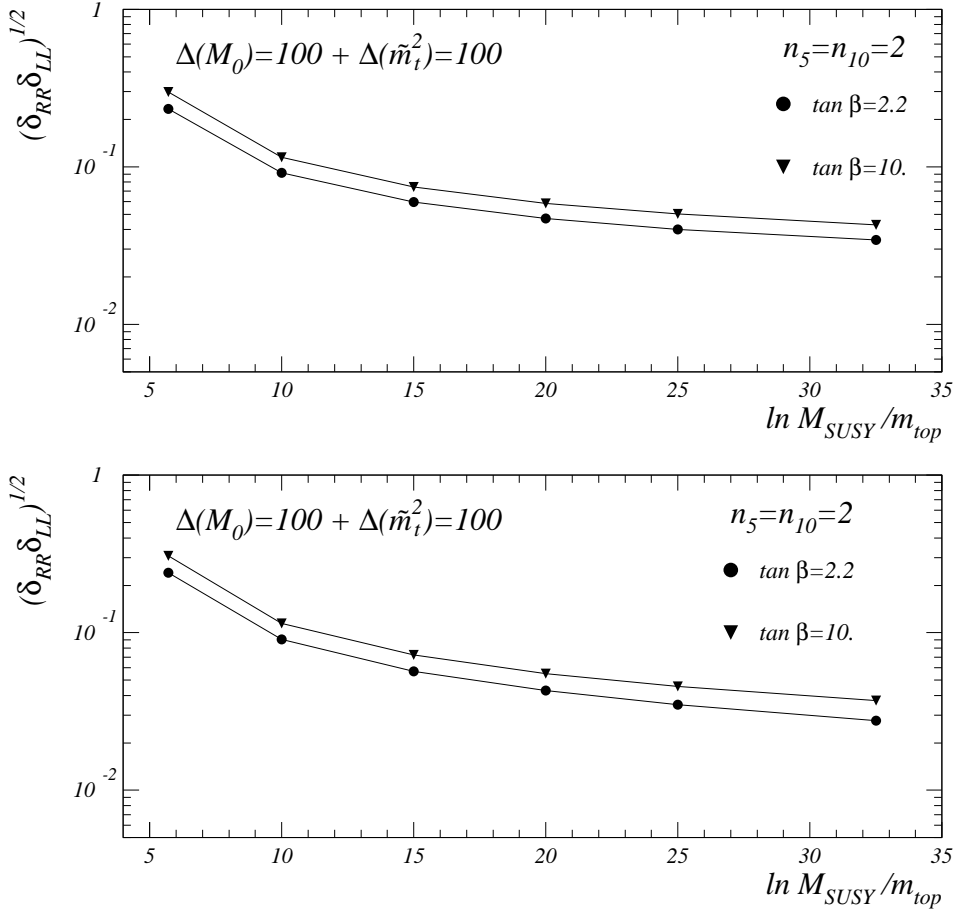}}
\caption{Maximum value for $(\delta_{LL}\delta_{RR})^{1/2}$
that is consistent with $\Delta(m^2_Z, M_{3,0}) <100$, 
$\Delta(m^2_Z, m^2_{\tilde{t},0}) <100$ and $(\Delta m_K)_{SUSY} 
<(\Delta m_K)_{exp}$. Two boundary conditions
are considered: $m^2_{H_u,0}=0$ (top) and gauge-mediated relations
(bottom). Two values for $\tan \beta$ are considered.
The value of $A_{t,0}$ was chosen to maximize the value of
the stop masses at the weak scale.}
\protect\label{m2ftd1}
\end{figure}

We show the limits on $\sqrt{\delta_{LL} \delta_{RR}}$
as a function of $M_{SUSY}$ for the case
$(n_5 = 2, n_{10}= 2)$ in Figures \ref{m2ftd1}. 
In the top of Figure \ref{m2ftd1},
$m_{H_u,0}^2 = 0$ is assumed. GMSB relations between the
stop and Higgs masses are assumed in the bottom of Figure \ref{m2ftd1}. 
For both cases, $\Delta _{max}=100$, $\tan \beta = 2.2$ and $10$
are considered. For other choices for $\Delta _{max}$, the 
upper limit to $\delta$ roughly scales as $\sqrt{\Delta _{max}/100}$, since
both $m_{\tilde{t},max}$, $M_{3,max}$ and therefore $M_{S,max}$
scale as $\sqrt{\Delta _{max}}$.

In the case of GMSB mass relations, the boundary value of the
Higgs mass and the stop masses
are comparable for high $M_{SUSY}$.
Since $m_{H_u,0}^2$ results in a negative
contributon to the stop mass squared,
this tends to reduce
the stop mass squared at the weak scale as
compared to the case $m_{H_u,0}^2 = 0$.
Then, from the above analysis, we can see that $M_{S,max}$ and,
in turn, the limit on $\delta$ is smaller for the GMSB case as
compared to the case $m_{H_u,0}^2 = 0$. This can be seen by comparing the 
top and bottom of Figure \ref{m2ftd1}.

\begin{figure}
\centerline{\epsfxsize=0.8\textwidth \epsfbox{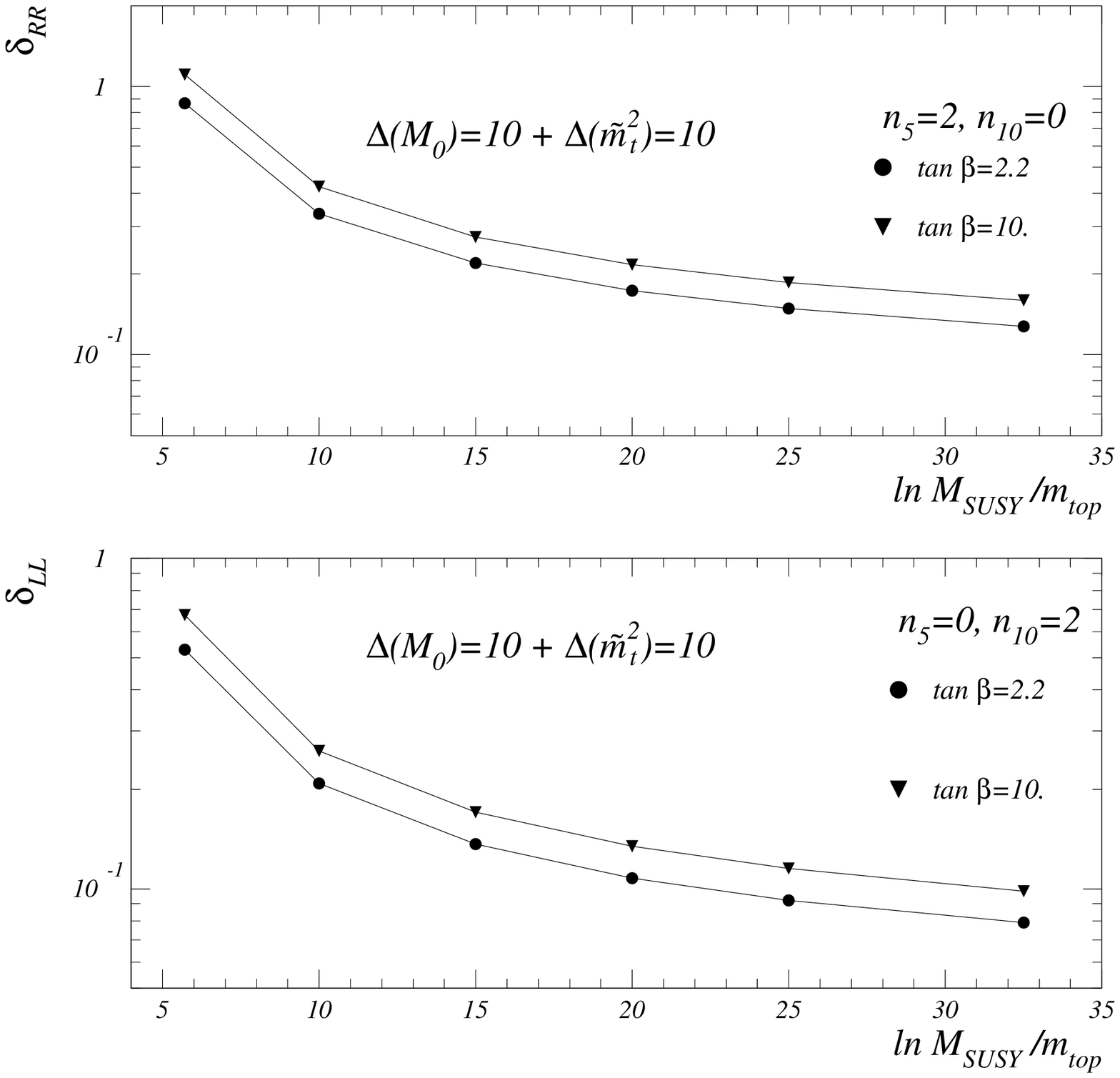}}
\caption{Maximum value for $\delta_{LL}$, $\delta_{RR}$
that is consistent with $\Delta(m^2_Z, M_{3,0}) <10$,
$\Delta(m^2_Z, m^2_{\tilde{t},0}) <10$ and $(\Delta m_K)_{SUSY}                <(\Delta m_K)_{exp}$. It was assumed that
$m^2 _{H_u,0}=0$. Two values for $\tan \beta$ are considered.
The value of $A_{t,0}$ was chosen to maximize the value of
the stop masses at the weak scale.}
\protect\label{m2ftd2}
\end{figure}

In Figure \ref{m2ftd2} the limits on
$\delta_{RR}$ and $\delta_{LL}$ for $(n_5 = 2, n_{10}= 0)$
and $(n_5 = 0, n_{10}= 2)$ are shown, respectively. We assume
$m_{H_u,0}^2 = 0$ and consider $\tan \beta = 2.2$ and $10$.
If we choose $\Delta _{max}$ to be 100, then
we get a constraint on $\delta$ ($\delta \stackrel{<}{\sim} 0.5$)
only for high
values of $M_{SUSY}$.
So, we choose instead $\Delta _{max}$ to be 10.

We have checked that,
for $\tan \beta = 10$, the limits on the boundary
value of the stop mass from requring positivity of the mass squared
at the weak scale do not differ by more than a few percent from the
case $\tan \beta =2.2$ (for the same values of the gluino and heavy scalar
masses).
However, the fine tuning of EWSB for the same gluino and stop mass
is smaller for $\tan \beta = 10$ as compared to $\tan \beta = 2.2$.
This is because,
for $\tan \beta = 10$, $\lambda_t$ is smaller than in the case
$\tan \beta =2.2$. Hence the
sensitivity of the weak scale value of $m_{H_u}^2$ to $m^2_{\tilde{t},0}$ and
$M_{3,0}$ is smaller. Also, the $\tan^2 \beta / (\tan^2 \beta -1)$
factor in Equation \ref{muEW} is smaller, 
further reducing the sensitivity of $m^2_Z$ to
$m^2_{\tilde{t},0}$ and
$M_{3,0}$. In other words, for $\tan \beta = 10$,
$m_{\tilde{t},max}$ and $M_{3,max}$ are larger so that
$M_{S,max}$ and, in turn, the limit on $\delta$ is larger.
This can be seen in Figures \ref{m2ftd1} and \ref{m2ftd2}.

\section{Conclusions}

In this paper we have studied whether the SUSY flavor problem
can be solved by
making the scalars of the first and second
generations heavy, with masses $M_S$
($\stackrel{>} \sim
$few TeV), without destabilising the weak scale. If the
scale, $M_{SUSY}$, at which SUSY breaking is mediated to the SM scalars
is close to the GUT scale, then the heavy scalars
drive the light scalar (in particular the stop) mass
squareds
negative through two-loop RG evolution. In order
to keep the mass squareds at the weak scale positive,
the initial value
of the stop (and other light scalar) soft masses, $m_{\tilde{f}_i,0}$,
must typically be
$\stackrel{>} \sim 1$ TeV, leading
to fine tuning in EWSB. We included two new effects in this
analysis:
the effect of $\lambda _t$ in the RGEs which makes
the stop mass
squareds at the weak scale more negative and hence makes
the constraint                           
on the initial value stronger, and the QCD corrections
to the
SUSY box diagrams which contribute to $K-\bar{K}$
mixing.

Some results of our analysis for $M_{SUSY} = M_{GUT}$
can be summarized as follows. We restrict the gluino
mass (at the weak
scale) to be less than about $710$ GeV, so that the fine
tuning of
$m^2_Z$ with respect to the bare gluino mass, $M_{3,0}$, is not worse
than $1\%$. This requires that $M_{3,0} \ltap 300$ GeV.
We also assume that $m_{H_u ,0}^2 = 0$ to 
maximize the value of the stop masses at the weak scale.
We find that 
if $\sqrt{\delta_{LL} \delta_{RR}}=0.22$ then $M_S \geq$ 40 TeV 
is required to be consistent with $\Delta m_K$.  With
these assumptions, 
this implies that 
for $M_{3,0}$ less than 1 TeV, 
$m_{\tilde{t}_i,0} >$ 6.5 TeV is needed 
to not break colour and charge at the 
weak scale.
Even for $\sqrt{\delta _{LL} \delta _{RR}} = 0.04$,
we find that
we need $M_S \stackrel{>} \sim 7$ TeV.
This implies that $m_{\tilde{t},0} >1$ TeV is required 
if $M_{3,0}\leq $ 300 GeV. 
This results in
a fine tuning of $\sim 1\%$.        
For $\delta_{LL} = 1$ and $\delta _{RR} = 0$,
we find that $M_S \stackrel{>} \sim 30$ TeV and
$m_{\tilde{t},0} > 4.5$ TeV.
For $\delta_{LL} = 0.22$ and $\delta _{RR} = 0$,
we find that $M_S \stackrel{>} \sim 7$ TeV and
$m_{\tilde{t},0} > 1$ TeV.
For $\delta_{LL} = 0$ and $\delta _{RR} = 1$,
we find that $M_S \stackrel{>} \sim 30$ TeV and
$m_{\tilde{t}^c,0} > 2.5$ TeV.
The constraints are weaker for smaller values of $\delta$.
In a realistic model,
$m_{H_u ,0}^2$
might be comparable to $m_{\tilde{t},0}^2$ and the
constraints
on $m_{\tilde{t},0}$ in this case 
are stronger. This is also discussed.
We note that
independent of the constraint from
$K-\bar{K}$ mixing, our analysis can be used to check the
phenomenological viability of any model that has
heavy scalars. We also discuss the phenomenological viability of 
the anomalous $D-$term solution, and find it to be problematic.

We then considered the possibility that $M_{SUSY} = M_S$.
In this case, there is no RG log enhancement of the
negative contribution of the heavy scalar masses to
the light scalar masses.
For this case, we computed the   
finite parts of the two-loop diagrams
and used these results as estimates of the two-loop
contribution
of the heavy scalars to the light scalar soft mass squareds.
We then combined these results with the constraints
from $K-\bar{K}$ mixing
to obtain lower limits on the boundary values of the stops.
As an example, we assumed gauge mediated SUSY breaking boundary conditions
for the light scalars. If $n_5 \neq n_{10}$ then one of the 
selectron masses, 
rather than the stop masses, provides the stronger constraint 
on $m_{\tilde{t}_i,0}$ once gauge-mediated boundary 
conditions are used to relate $m_{\tilde{e}^c,0}$ and $m_{\tilde{L},0}$
to   
$m_{\tilde{t}_i,0}$. Some of our results
can be summarized as follows. We restrict the gluino
mass at the weak scale
to be less than about $2.3$ TeV, again
to avoid more than $1\%$ fine tuning of $m^2_Z$
with respect to the gluino mass. 
For $\sqrt{\delta_{LL} \delta_{RR}}=.22$ we find that 
$m_{\tilde{t}_i,0} \geq$ 1.4 TeV is required. 
The fine tuning of $m^2_Z$ with respect to the stop
mass is $\sim 1.5\%$ in this case.
For the cases $\delta_{LL} = 0$ and $\delta _{RR} = 1$,
and $\delta_{LL} = 1$
and $\delta _{RR} = 0$ we find that $m_{\tilde{t},0}
\stackrel{>} \sim 1$ TeV.
As before, the constraints on $m_{\tilde{t},0}$ for smaller
values of
$\delta$ are weaker than $\sim 1$ TeV. Again, we emphasize 
that the constraints in an 
actual model of this low-energy supersymmetry breaking scenario 
could be different, and our results should
be treated as estimates only. We also briefly discuss 
the  
$CP$ violating constraints from $\epsilon$, and find that 
these limits increase
by a factor of $\sim 12$ if $O(1)$ phases are present. 
 
Finally, in Section 5 the scale of
supersymmetry breaking is varied between 50 TeV and
$2 \times 10^{16}$ GeV. Uppers bounds to $\delta$, 
that are  
consistent with positivity of the light scalar masses, naturalness
in electroweak symmetry breaking, and $(\Delta m_K)_{exp}$, are obtained.
These results are summarized in Figures \ref{m2ftd1} and \ref{m2ftd2}.

\section{Acknowledgements}
The authors would like to acknowledge N. Arkani-Hamed,
I. Hinchliffe, H. Murayama and
M. Suzuki for suggestions and discussions. The authors also 
thank B. Nelson for drawing our attention to the omission of 
the one-loop hypercharge $D$-term in an earlier version of the 
manuscript. 
This work was supported in part by the Director, Office of
Energy Research, Office of High Energy and Nuclear Physics,
Division of
High Energy Physics of the
U.S. Department of Energy under Contract
DE-AC03-76SF00098. 
KA was also supported
by the Berkeley Graduate Fellowship and
MG by NSERC.

\section{Appendix: Two-loop calculation}
\begin{figure}
\centerline{\epsfxsize=0.49\textwidth \epsfbox{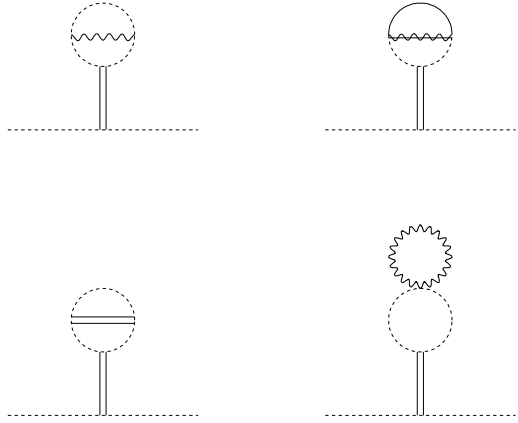}}
\caption{Mixed two-loop corrections to the scalar mass. Wavy lines, 
wavy lines with a straight line
through them, solid lines, and dashed lines denote gauge boson, 
gaugino, fermion and scalar propagators,respectively. 
The double-line denotes the
hypercharge $D$-term propagator.}
\protect\label{mixedtwoloopdiag}
\end{figure}

In this Appendix we discuss the two-loop contribution of the 
heavy scalar soft masses to the
light scalar soft masses. These contributions can be divided into 
two classes. In the first class, a vev for the hypercharge $D$-term 
is generated at two-loops. The Feynman diagrams for these contributions 
are given in Figure \ref{mixedtwoloopdiag} and 
are clearly $\sim \alpha_1 \alpha_i$.
These diagrams are computed in a later portion of this Appendix. 
In the other class, the two-loop diagrams are $\sim \alpha^2_i$. 
These have been
computed by Poppitz and Trivedi\cite{poppitz}. So, we will 
not give details of this
computation which can be found in their paper. However, 
our result for the
finite parts of these diagrams differs slightly from theirs 
and we discuss the
reason for the discrepancy. When one regulates the theory 
using dimensional
reduction \cite{dred,epscalar} (compactifying to $D < 4$ dimensions), the 
vector field decomposes
into a $D$-dimensional vector and $4-D$ scalars, 
called $\epsilon$-scalars, in
the adjoint representation of the gauge group. 
Thus the number of Bose and Fermi degrees of freedom 
in the vector multiplet remain equal.
The $\epsilon$-scalars receive,
at one-loop, a divergent contribution to their mass, 
proportional to the
supertrace of the mass matrix of the matter fields. 
Neglecting the fermion masses,
this contribution is
\begin{equation}
\delta m^2_{\epsilon} = - \frac{\alpha}{4 \pi}
(\frac{2}{\epsilon}+\ln 4\pi-\gamma) (n_5 + 3 n_{10}) M_S^2.
\end{equation}
In our notation $D=4-\epsilon$.
Poppitz and Trivedi choose
the counterterm to cancel this divergence in the 
$MS$ scheme, {\it i.e.}, the
counterterm consists only of the divergent part, 
proportional to $1/\epsilon$.
When this counterterm is inserted in a one-loop 
$\epsilon$-scalar graph with SM
fields (scalars) as the external lines
, one obtains a divergent contribution to the
SM scalar soft masses (the $1/\epsilon$ of the counterterm 
is cancelled after
summing over the $\epsilon$ adjoint scalars running 
in the loop). Poppitz and
Trivedi use a cut-off, $\Lambda _{UV}$, to regulate this 
graph, giving a
contribution from this graph that is:
\begin{equation}
m^2_i = - \sum_{A} (n_5 + 3 n_{10})
C^i_A \frac{1}{16} (\frac{\alpha _A}{\pi})^2
M_S^2 \hbox{ln} \Lambda ^2 _{UV}
\end{equation}
with no finite part. We, on the other hand, 
choose the $\epsilon$-scalar mass
counterterm in the $\overline{MS}$ scheme, {\it i.e.}, 
proportional to $2/\epsilon -
\gamma + \hbox{ln}4 \pi$ (where $\gamma \approx 0.58$ 
is the Euler constant) and
use dimensional reduction to regulate the graph with the 
insertion of the
counterterm. This gives a contribution
\begin{eqnarray}
m^2_i & = & - \sum_{A} (n_5 + 3 n_{10})
C^i_A \frac{1}{16} \left(\frac{\alpha _A}{\pi}\right)^2
M_S^2 (\frac{2}{\epsilon} -
\gamma + \hbox{ln}4 \pi)^2 \epsilon \nonumber \\
 & = & - \sum _{A} (n_5 + 3 n_{10})
C^i_A \frac{1}{8} \left(\frac{\alpha _A}{\pi}\right)^2
M_S^2 (2/\epsilon -
2 \gamma + 2 \hbox{ln}4 \pi)
\end{eqnarray}
In the first line the 
first factor of $(2/\epsilon -
\gamma + \hbox{ln}4 \pi)$ is from the counter-term 
insertion, the second factor 
is the result of the loop integral, and 
the over-all factor of $\epsilon$ counts the number of 
$\epsilon$-scalars running in the loop. 
In the $\overline{MS}$ scheme, {\it i.e.}, after
subtracting $2/\epsilon -
\gamma + \hbox{ln}4 \pi$, we are left with a finite 
part\footnote{The same finite part is obtained in the $MS$ 
scheme,regulated with ${DR}^{\prime}$ .} proportional to $-
\gamma + \hbox{ln}4 \pi$. The remaining diagrams together 
give a finite result
and we agree with Poppitz and
Trivedi on this computation. Our result for the finite part
of the two-loop diagrams (neglecting the fermion masses) is
\begin{eqnarray}
m^2_{i,finite} (\mu) & = & -\frac{1}{8} 
\left( \hbox{ln} (4 \pi) - \gamma +
\frac{\pi^2}{3} - 2 - \hbox{ln} \left(\frac{M^2_S}{\mu ^2}\right) \right) 
\nonumber \\
 & & \times \sum_{A} \left( \frac{\alpha _A(\mu)}
{\pi} \right) ^2 (n_5 +
 3 n_{10}) C^i_A M^2_S
\end{eqnarray}
whereas the Poppitz-Trivedi result does not have 
the $\hbox{ln} (4 \pi) -
\gamma$ in the above result. The computation of the 
two-loop hypercharge
$D$-term, which gives contribution to the soft scalar 
mass squareds proportional
to $\alpha _1 \alpha _s$ and $\alpha _1 \alpha _2$ 
({\it i.e.}, the "mixed''
two-loop contributon) is discussed below in detail.

{\it Two-loop hypercharge D-term}

We compute the two-loop diagrams of
Figure \ref{mixedtwoloopdiag} in the Feynman gauge and set all 
fermion and gaugino masses to zero. It is 
convienent to calculate in this gauge 
because both the
scalar self-energy and the $D_Y$-term
vertex corrections are finite at one-loop 
and thus require no 
counter-terms. We have also computed the
two-loop diagrams in the Landau gauge and have found that 
it agrees with the calculation in the Feynman gauge. 
The calculation in the Landau gauge 
requires counter-terms, 
is more involved, and hence
the discussion is not included. 
Finally, 
in the calculation a global $SU(5)$ symmetry is 
assumed so that a hypercharge $D$-term is not 
generated at one-loop \cite{dimopoulos,nelson2}. 

The sum of the four Feynman diagrams in Figure \ref{mixedtwoloopdiag} 
is given 
in the Feynman gauge by
\begin{equation}
-i \tilde{\Pi}_{D,f}=i \frac{3}{5}
g^2_1 Y_f \hbox{Tr} Y_i\sum_{A,i} 
g_A^2 C_A^i(4 I_1(m^2_i)-4I_2(m^2_i)+I_3(m^2_i))
\end{equation}
where the trace is over the gauge and flavour states of the 
particles in the loops. If the particles in the loop 
form complete $\bar{5}$ and $10$ representations with 
a common mass $M_S$, the 
sum simplifies to
\begin{equation}
-i \tilde{\Pi}_{D,f}=i \frac{3}{5}
\alpha_1 Y_f (n_5-n_{10})(\frac{4}{3} \alpha_3-\frac{3}{4} \alpha_2
-\frac{1}{12} \alpha_1) 
(4 I_1(M^2_S)-4I_2(M^2_S)+I_3(M^2_S)).
\end{equation}    

The functions $I_1$, $I_2$ and $I_3$
are
\begin{eqnarray}
I_1(m^2) & = & \int \frac{d^D p}{(2\pi)^D} 
\int \frac{d^D k}{(2\pi)^D}\frac{1}{(p^2-m^2)^2}
\frac{(2p-k)^2}{k^2}\frac{1}{(p-k)^2-m^2}, \\
I_2(m^2) & = & \int \frac{d^D p}{(2\pi)^D}
\int \frac{d^D k}{(2\pi)^D}\frac{1}{(p^2-m^2)^2}
\frac{k^2-k\cdot p}{k^2}\frac{1}{(p-k)^2}, \\
I_3(m^2) & = & \int \frac{d^D k}{(2\pi)^D}
\frac{1}{(k^2-m^2)^2}
\int \frac{d^D q}{(2\pi)^D}\frac{1}{q^2-m^2}.
\end{eqnarray}

We now compute these functions.

{\it \underline{Evaluating $I_1$}}

After a Feynman parameterization and performing a
change of variables, $I_1=J_1+J_2$, where
\begin{equation}
J_1(m^2)=\Gamma(3)\int_{0}^{1} dx (1-x) \int \frac{d^D p}{(2\pi)^D}
\int \frac{d^D k}{(2\pi)^D} 4 \frac{p^2}{k^2}
\frac{1}{(p^2-(m^2-x(1-x)k^2))^3}
\end{equation}
and 
\begin{equation}
J_2(m^2)=\Gamma(3)\int_{0}^{1} dx (1-x) (2x-1)^2
 \int \frac{d^D p}{(2\pi)^D}
\int \frac{d^D k}{(2\pi)^D} 
\frac{1}{(p^2-(m^2-x(1-x)k^2))^3}.
\end{equation}


After some algebra we find that 
\begin{equation}
J_1(m^2)=\frac{\Gamma(3-D)}{(4\pi)^D}(m^2)^{D-3}
\frac{2 D}{D/2-1} B(2-D/2,3-D/2),
\end{equation}
\begin{equation}
J_2(m^2)
 =  \frac{\Gamma(3-D)}{(4\pi)^D}(m^2)^{D-3}
\times
(4 B(3-D/2,2-D/2)-4 B(2-D/2,2-D/2) 
 +B(1-D/2,2-D/2)), 
\end{equation}
where $B(p,q)=\Gamma[p] \Gamma[q]/ \Gamma[p+q]$ is the usual 
Beta function.

Combining these two results gives
\begin{equation}
I_1(m^2)
=\frac{\Gamma(3-D)}{(4\pi)^D} (m^2)^{D-3}
\frac{1-D}{D-2}B(3-D/2,2-D/2).
\end{equation}

{\it \underline{Evaluating $I_2$}}

\begin{eqnarray*}
I_2(m^2)&=&\int \frac{d^D p}{(2\pi)^D}
\int \frac{d^D k}{(2\pi)^D}\frac{1}{(p^2-m^2)^2}
\frac{k^2-k\cdot p}{k^2}\frac{1}{(p-k)^2} \\
& = & \frac{1}{(4 \pi)^D} 
\Gamma(3-D) (m^2)^{D-3} B(D/2,1-D/2).
\end{eqnarray*}

{\it \underline{Evaluating $I_3$}}

\begin{eqnarray*}
I_3(m^2) & = & \int \frac{d^D k}{(2\pi)^D} 
\frac{1}{(k^2-m^2)^2}
\int \frac{d^D q}{(2\pi)^D}\frac{1}{q^2-m^2} 
 \\
& = & \left(\frac{i}{(4\pi)^{D/2}} \Gamma(2-D/2) 
(m^2)^{D/2-2}\right)
\left(\frac{i}{(4\pi)^{D/2}}\frac{\Gamma(2-D/2)}
{D/2-1} (m^2)^{D/2-1}\right) \\
& = & -\frac{1}{(4\pi)^D}(\Gamma(2-D/2))^2 
\frac{1}{D/2-1}(m^2)^{D-3}.
\end{eqnarray*}

We may now combine $I_1$, $I_2$ and $I_3$ to obtain
\begin{eqnarray*}
T(m^2) & \equiv &
4I_1(m^2)-4I_2(m^2)+I_3(m^2) \\
& = & \frac{(m^2)^{D-3}}{(4 \pi)^D} \times (
4\left(\frac{1-D}{D-2}B(3-D/2,2-D/2)-B(D/2,1-D/2)\right) 
\Gamma(3-D) \\
& & -\frac{1}{D/2-1} \Gamma(2-D/2)^2 ).
\end{eqnarray*}
Writing $D=4-\epsilon$ and expanding in $\epsilon$ gives 
\begin{equation}
T(m^2)=\frac{1}{(16 \pi^2)^2}\left(\frac{4}{\epsilon}
+\left(6-\frac{2}{3} \pi^2+4(\ln (4 \pi)-\gamma)-4 \ln m^2\right) m^2
+O(\epsilon)\right).
\end{equation}
In the $\overline{MS}$ scheme the combination 
$2/\epsilon+\ln(4 \pi)-\gamma$ is subtracted 
out. The finite piece that remains is 
\begin{equation}
\frac{1}{(16 \pi^2)^2}
\left(6-\frac{2}{3} \pi^2+2(\ln (4 \pi)-\gamma)-4 \ln m^2\right) m^2.
\end{equation}
Thus in the $\overline{MS}$ scheme 
\begin{eqnarray}
-i\tilde{\Pi}_{D,f} & = & i \frac{3}{5}\frac{1}{(16 \pi^2)^2}
 g^2_1Y_f \hbox{Tr}_i Y_i\sum_A 
g_A^2 C^i_A \left(6-\frac{2}{3} \pi^2
+2(\ln (4 \pi)-\gamma)-4 \ln m_i^2\right) m_i^2 \nonumber \\
 & = & i \frac{3}{5}\frac{1}{16 {\pi}^2}
 \alpha_1(\mu_R)(n_5-n_{10}) Y_f
\left(6-\frac{2}{3} {\pi}^2+2 (\ln (4 \pi)-\gamma)
-4 \ln(\frac{M_S^2}{\mu_R^2}) \right) \nonumber \\
 & & \times
\left(\frac{4}{3} \alpha_3(\mu_R)-\frac{3}{4}
\alpha_2(\mu_R)-\frac{1}{12}\alpha_1(\mu_R)\right) M_S^2 
\end{eqnarray}

\end{document}